\RequirePackage{fix-cm}

\documentclass[twocolumn,epjc3]{svjour3} 

\smartqed  

\RequirePackage{graphicx}

\RequirePackage{siunitx}
\RequirePackage[version=4]{mhchem}
\RequirePackage{booktabs}
\RequirePackage{multirow}
\RequirePackage[colorlinks,citecolor=blue,urlcolor=blue,linkcolor=blue]{hyperref}
\RequirePackage{dblfloatfix}
\RequirePackage{microtype}

\newcommand{\nuebar}{\overline{\nu}_e}

\newcommand{\avg}[1]{\left< #1 \right>}
\newcommand{\kforty}{\ce{ ^{40}K }}

\DeclareSIUnit\solarmass{\ensuremath{M_\odot}}
\DeclareSIUnit\parsec{\ensuremath{pc}}

\journalname{Eur. Phys. J. C}

\begin{document}

\title{The KM3NeT potential for the next core-collapse supernova observation with neutrinos
}
\subtitle{The KM3NeT Collaboration}


\onecolumn

\begin{NoHyper} 

\author{
S.~Aiello\thanksref{a}
\and
A.~Albert\thanksref{b,bd}
\and
S. Alves Garre\thanksref{c}
\and
Z.~Aly\thanksref{d}
\and
A. Ambrosone\thanksref{e,f}
\and
F.~Ameli\thanksref{g}
\and
M.~Andre\thanksref{h}
\and
G.~Androulakis\thanksref{i}
\and
M.~Anghinolfi\thanksref{j}
\and
M.~Anguita\thanksref{k}
% ...
\and
(the KM3NeT Collaboration)
}
\institute{
\label{a}INFN, Sezione di Catania, Via Santa Sofia 64, Catania, 95123 Italy
\and
\label{b}Universit{\'e}~de~Strasbourg,~CNRS,~IPHC~UMR~7178,~F-67000~Strasbourg,~France
\and
\label{c}IFIC - Instituto de F{\'\i}sica Corpuscular (CSIC - Universitat de Val{\`e}ncia), c/Catedr{\'a}tico Jos{\'e} Beltr{\'a}n, 2, 46980 Paterna, Valencia, Spain
\and
\label{d}Aix~Marseille~Univ,~CNRS/IN2P3,~CPPM,~Marseille,~France
\and
\label{e}INFN, Sezione di Napoli, Complesso Universitario di Monte S. Angelo, Via Cintia ed. G, Napoli, 80126 Italy
\and
\label{f}Universit{\`a} di Napoli ``Federico II'', Dip. Scienze Fisiche ``E. Pancini'', Complesso Universitario di Monte S. Angelo, Via Cintia ed. G, Napoli, 80126 Italy
\and
\label{g}INFN, Sezione di Roma, Piazzale Aldo Moro 2, Roma, 00185 Italy
\and
\label{h}Universitat Polit{\`e}cnica de Catalunya, Laboratori d'Aplicacions Bioac{\'u}stiques, Centre Tecnol{\`o}gic de Vilanova i la Geltr{\'u}, Avda. Rambla Exposici{\'o}, s/n, Vilanova i la Geltr{\'u}, 08800 Spain
\and
\label{i}NCSR Demokritos, Institute of Nuclear and Particle Physics, Ag. Paraskevi Attikis, Athens, 15310 Greece
\and
\label{j}INFN, Sezione di Genova, Via Dodecaneso 33, Genova, 16146 Italy
\and
\label{k}University of Granada, Dept.~of Computer Architecture and Technology/CITIC, 18071 Granada, Spain
% ...
\and
\label{bd}Universit{\'e} de Strasbourg, Universit{\'e} de Haute Alsace, GRPHE, 34, Rue du Grillenbreit, Colmar, 68008 France
% ...
} 
\thankstext{corr1}{e-mail: mcolomer@apc.in2p3.fr}
\thankstext{corr2}{e-mail: lincetto@cppm.in2p3.fr}

\date{Received: date / Accepted: date}

\maketitle

\end{NoHyper} 

\twocolumn

\begin{abstract}
The KM3NeT research infrastructure is under construction in the Mediterranean Sea. It consists of two water Cherenkov neutrino detectors, ARCA and ORCA, aimed at neutrino astrophysics and oscillation research, respectively. Instrumenting a large volume of sea water with $\sim\num{6200}$ optical modules comprising a total of $\sim \num{200000}$ photomultiplier tubes, KM3NeT will achieve sensitivity to $\sim\SI{10}{MeV}$ neutrinos from Galactic and near-Galactic core-collapse supernovae through the observation of coincident hits in photomultipliers above the background. In this paper, the sensitivity of KM3NeT to a supernova explosion is estimated from detailed analyses of background data from the first KM3NeT detection units and simulations of the neutrino signal. The KM3NeT observational horizon (for a $5\,\sigma$ discovery) covers essentially the Milky-Way and for the most optimistic model, extends to the Small Magellanic Cloud ($\sim \SI{60}{\kilo\parsec}$). Detailed studies of the time profile of the neutrino signal allow assessment of the KM3NeT capability to determine the arrival time of the neutrino burst with a few milliseconds precision for sources up to \SIrange{5}{8}{\kilo\parsec} away, and detecting the peculiar signature of the \textit{standing accretion shock instability} if the core-collapse supernova explosion happens closer than \SIrange{3}{5}{\kilo\parsec}, depending on the progenitor mass. KM3NeT's capability to measure the neutrino flux spectral parameters is also presented.

\end{abstract}

\section{Introduction}
    Core-collapse supernovae (CCSNe) are explosive phenomena that may occur at the end of the life of massive stars. In a typical CCSN, an amount of energy as large as $\SI{3E53}{erg}$ can be released mainly through the emission of a burst of neutrinos having a mean energy in the \SIrange{10}{20}{MeV} range. Neutrinos carry $\sim99\%$ of the progenitor's gravitational energy and are believed to play an important role in the explosion mechanism. The neutrino burst is emitted on a timescale of about ten seconds from the onset of the collapse. At this stage, the star envelope is opaque to the electromagnetic radiation. As a consequence, neutrino detection can occur a few hours before the supernova becomes visible to electromagnetic observatories. An overview of CCSN neutrino phenomenology is given in Refs.~\cite{Giunti:2007ry,Janka}. The first and only supernova neutrinos were observed from the SN 1987A explosion in the Large Magellanic Cloud. Two dozen events were detected by three neutrino detectors in operation at that time~\cite{Kamioka,IMB,Baksan}. With the new generation of neutrino detectors, the observation of the next CCSN will provide invaluable insights into the astro-, subnuclear and nuclear physics involved in these extreme phenomena.
    
    The KM3NeT neutrino detectors, ARCA and ORCA (\emph{Astrophysics} and \emph{Oscillation Research with Cosmics in the Abyss}), are under construction in the Mediterranean Sea~\cite{KM3NeT:2016-LoI}. They will instrument a volume of seawater on the $\si{km^3}$ scale with about $\num{200000}$ photomultiplier tubes (PMTs). Their primary goals are the detection of astrophysical TeV--PeV neutrinos and the precise measurement of the neutrino oscillation properties, respectively. The sensitivity to neutrinos at the $\SI{10}{MeV}$ scale can be achieved through the observation of a collective increase in the coincidence counting rates of the optical modules, exploiting their multi-PMT design.
    
    In this work, the KM3NeT sensitivity to a CCSN neutrino burst is presented. The CCSN mechanism and the flux models are introduced in Section~\ref{s:flux-models}. The KM3NeT detectors are described in Section~\ref{s:detector}. The detection method of the CCSN neutrino burst and the KM3NeT sensitivities are presented in Section~\ref{s:ccsn-km3net} and Section~\ref{s:sensitivity}, respectively. The systematic uncertainties affecting the detection capability are covered in Section~\ref{s:systematics}. The potential to resolve the mean neutrino energy is shown in Section~\ref{s:energy}. The analyses related to the neutrino burst time profile are introduced in Section~\ref{s:lc}. Two time-dependent analyses, evaluating the possibility to infer the arrival time of the signal and to observe hydrodynamical instabilities in the CCSN accretion phase, are described in Sections~\ref{s:t0} and~\ref{s:sasi}, respectively. 
    
    \section{Neutrinos from core-collapse supernov\ae}
    \label{s:flux-models}
    Progenitor stars with a mass above ten solar masses go through several nuclear fusion stages as they reach the end of their life cycle. In its final state, the star consists of an iron core surrounded by shells of lighter elements. The core is in hydrostatic equilibrium between the pressure of degenerate electrons and the gravitational force. The stellar evolution is driven by processes of iron photo-dissociation and electron capture: 
    
    \begin{equation}
        \gamma + \ce{^{56}Fe} \rightarrow 13\,\alpha + 4 \,\ce{n} \: ; 
    \end{equation}

    \begin{equation}
        e^{-} + p \rightarrow n + \nu_e \: .
    \end{equation}
    
    The two processes result in the progressive reduction of the density and average kinetic energy of electrons. At some point, the equilibrium is broken and the iron core collapses to form a proto-neutron star. The infalling matter bounces off the core, producing a shock wave. The shock propagates to the outer layers at a speed of $\sim\SI{E8}{m.s^{-1}}$, losing energy in the photo-dissociation of nuclei. Neutrinos produced in electron captures are confined behind the shock as long as it propagates through densities above $\sim\SI{E11}{g.cm^{-3}}$. At the crossing of this density threshold, a first pulse-like emission of electron neutrinos, called \emph{breakout} or \emph{neutronisation burst}, occurs. Due to the energy loss in the propagation, the shock wave eventually stalls. In the so-called \emph{accretion phase}, matter keeps falling through the stalled shock into the core. This induces a strong emission of neutrinos and anti-neutrinos dominated by the electronic flavour. At this stage, different processes can contribute to the so-called \textit{neutrino heating} that revives the shock and leads to the expulsion of the envelope. Hydrodynamical instabilities, convective motions in the mantle and acoustic oscillations of the neutron star are believed to play a role in determining the outcome of the explosion~\cite{Burrows_2007,obergaulinger_hammer_muller_2006,OConnor:2018tuw}. In the last phase, the core undergoes a thermal cool-down that can last up to tens of seconds. A review of the full process can be found in Ref.~\cite{Janka}.
    
    Following Ref.~\cite{Tamborra:2014hga-3D}, the CCSN neutrino energy spectrum can be described as a function of the neutrino energy, $E$, and the time, $t$, relative to the core bounce (at $t=0$) as:
    \begin{equation}
    \frac{d \Phi}{dE\,dt} (E, t) = \frac{L(t)}{4 \pi d^2}f(E, \avg{E(t)}, \alpha(t)) \: ,
    \label{eq:dphi-de}
    \end{equation}
    where $\avg{E} $ is the mean neutrino energy, $L$ the neutrino luminosity, $d$ the distance to the source, and $\alpha$ the spectral shape parameter. At a given time, the energy dependence of the spectrum follows a quasi-thermal distribution \cite{Keil:2003sw}:
    \begin{equation}
    f(E, \avg{E}, \alpha) = \frac{E^\alpha}{\Gamma(1 + \alpha)} \left ( \frac{1 + \alpha}{\avg{E}}\right )^{1 + \alpha} e^{\frac{-E (1 + \alpha)}{\avg{E}}} \: ,
    \end{equation}
    where $\Gamma$ is the Euler gamma function. The spectral shape parameter, $\alpha$, is defined as:
    \begin{equation}
    \alpha = \frac{ \avg{E^2} - 2\avg{E}^2 }{ \avg{E}^2 - \avg{E^{2}} }\: .
    \end{equation} 
    For $\alpha = 2$ the expression reduces to a Maxwell-Boltzmann distribution, while for $\alpha > 2$ the spectrum is pinched, i.e. it has smaller width and is peaked at higher energy.
    
    State of the art three-dimensional simulations of CCSNe predict the development of fast and asymmetric hydrodynamic motions in the core during the accretion phase \cite{Tamborra:2013laa-SASI}. In particular, the \emph{standing accretion shock instability} (SASI)~\cite{Tamborra:2013laa-SASI,Lund2010} phenomenon may produce oscillations of the core, reflected in the time profile of the neutrino emission (\emph{neutrino light curve}). This asymmetric instability is believed to favour the explosion by enhancing the neutrino energy deposition on the shock (\emph{neutrino heating}). Some models identify the SASI oscillation as a potential source of gravitational waves~\cite{Roma:2019kcd}.
    
    The supernova neutrino detection sensitivities presented in this paper are computed considering the fluxes predicted by 3D simulations from the Garching Group\footnote{https://wwwmpa.mpa-garching.mpg.de/ccsnarchive/}. The considered fluxes correspond to the cases of two CCSNe from progenitors with respective masses of $\SI{11}{\solarmass}$ and $\SI{27}{\solarmass}$ \cite{Tamborra:2014hga-3D,Tamborra:2013laa-SASI}, and a so-called \emph{failed} supernova with a progenitor of $\SI{40}{\solarmass}$ \cite{Walk:2019miz-3DBH} collapsing into a black hole. A fourth CCSN progenitor of $\SI{20}{\solarmass}$~\cite{Tamborra:2014hga-3D,Tamborra:2013laa-SASI}, with enhanced SASI oscillations, is used in the light curve studies (see Section~\ref{s:lc}). 
    
\begin{figure*}[!ht]
        \hspace{-0.3cm}\includegraphics[width=0.345\textwidth]{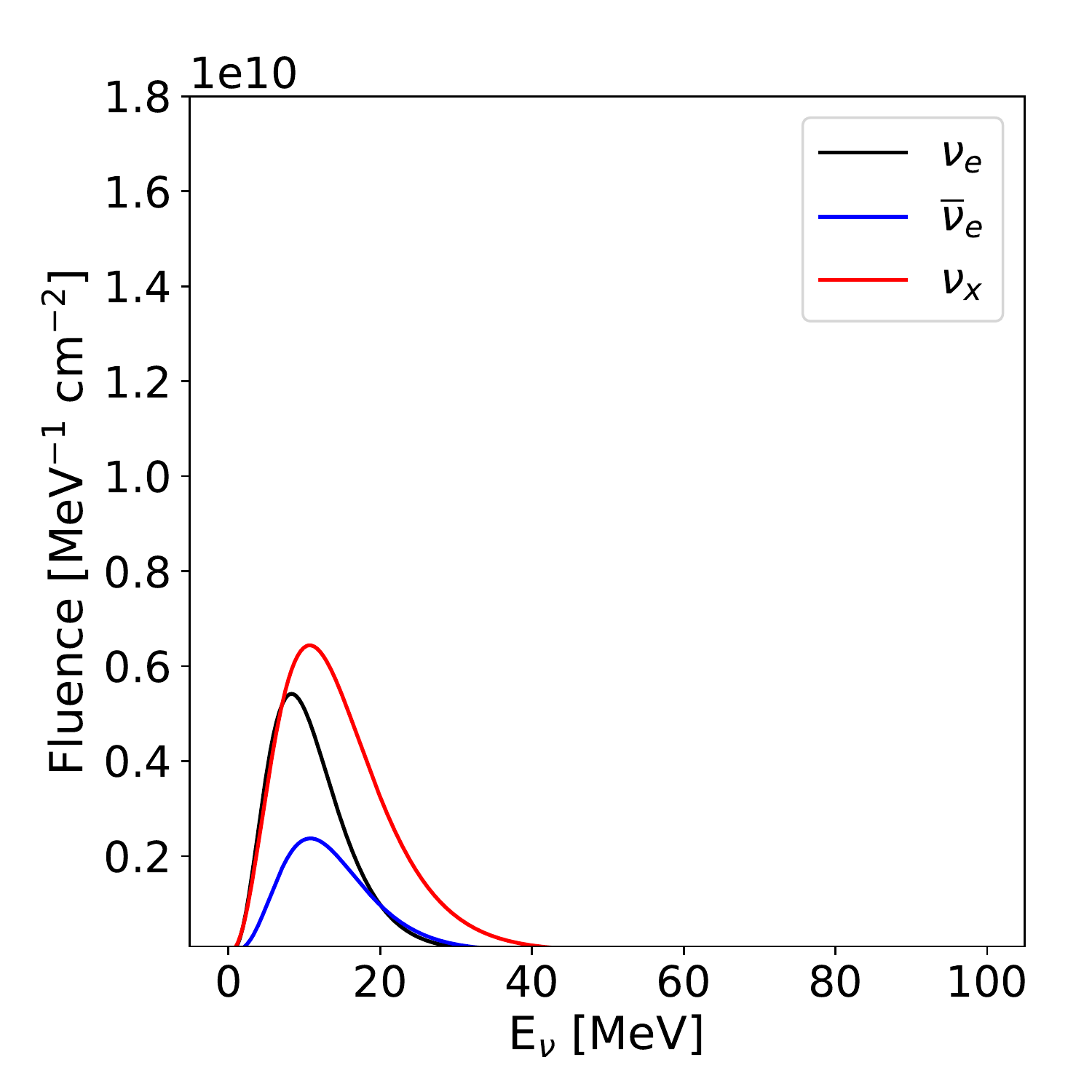}%
        \includegraphics[width=0.345\textwidth]{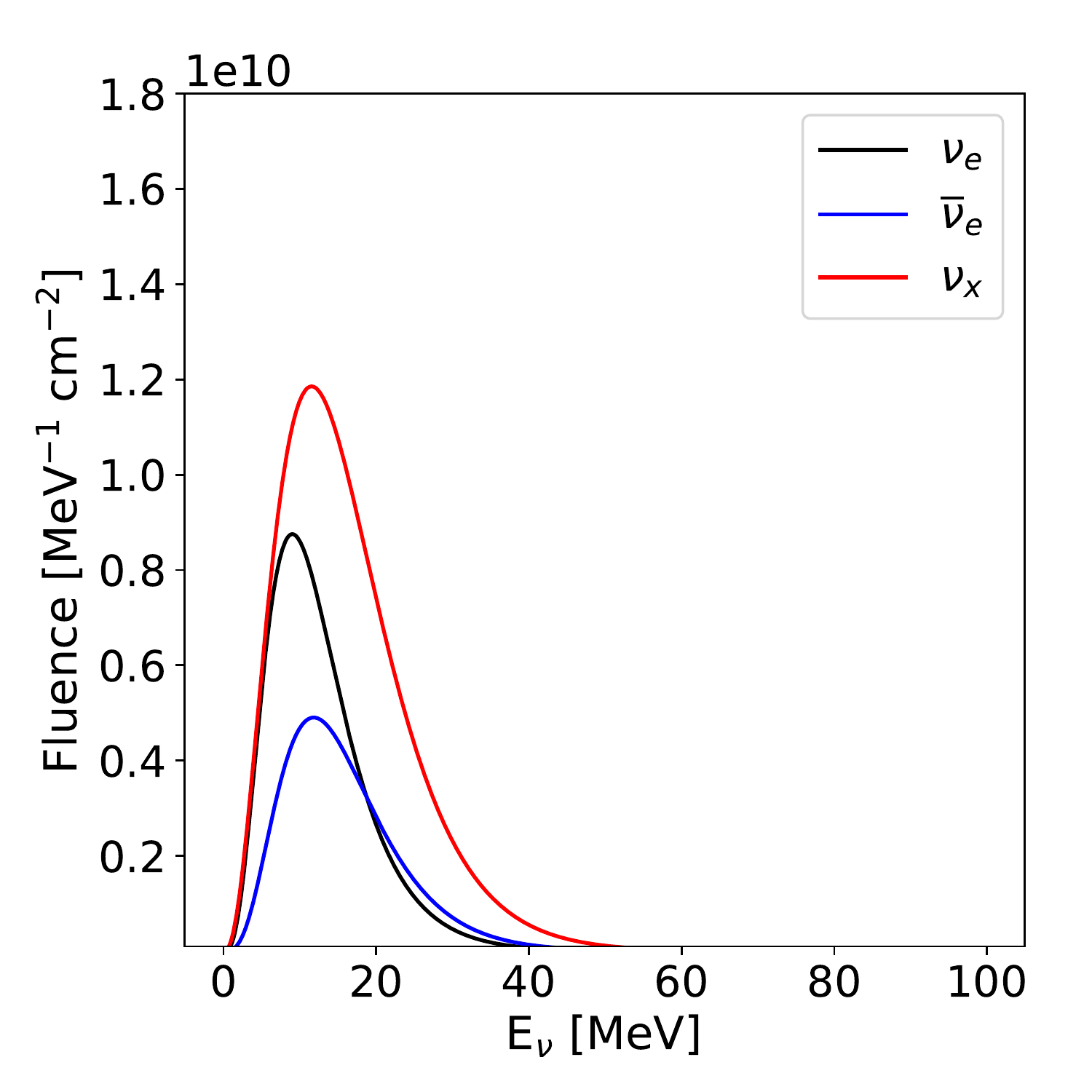}%
        \includegraphics[width=0.345\textwidth]{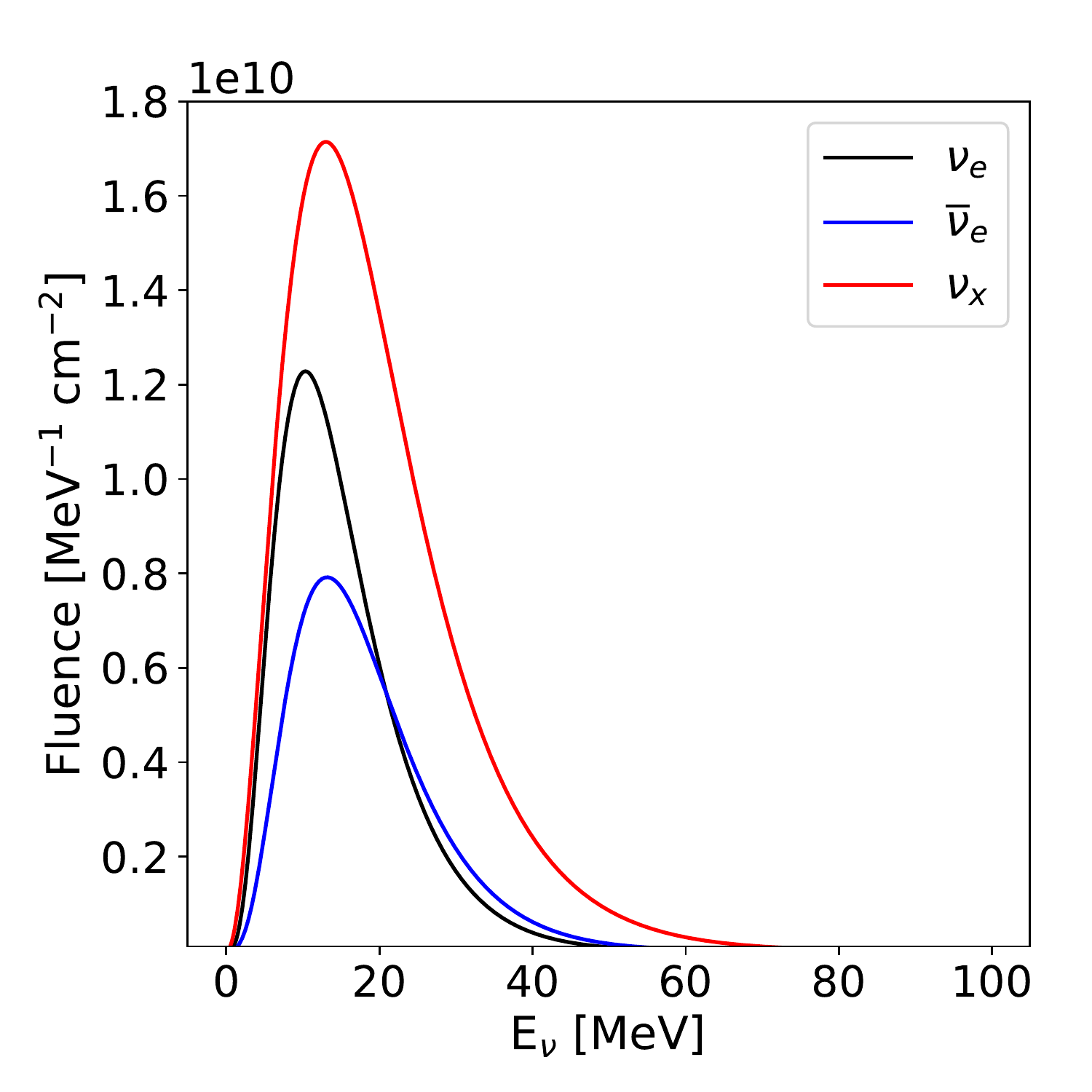}%
        \caption{Neutrino fluence during the accretion phase of a core-collapse supernova at $\SI{10}{kpc}$ from the Garching 3D simulation of three different progenitors with $\SI{11}{\solarmass}$ (left), $\SI{27}{\solarmass}$ (middle) and $\SI{40}{\solarmass}$ (right) over a time duration of $\SI{340}{ms}$, $\SI{543}{ms}$ and $\SI{562}{ms}$, respectively.}
        \label{fig:garching-fluence}
    \end{figure*}

    The simulated fluxes are centred on the accretion phase, including only the trailing edge of the neutronisation burst and stopping before the cooling phase. While these 3D simulations of the 11, 27 and $\SI{20}{\solarmass}$ progenitors do not reproduce the final explosion, they are here considered as reliable estimates of the neutrino flux emitted during the accretion. For the case of KM3NeT detectors, mainly sensitive to $\nuebar$ in the $\sim\SI{10}{MeV}$ energy range (see Section~\ref{s:ccsn-km3net}), the fraction of detected neutrinos from the unaccounted $\nu_e$ breakout pulse is less than $10^{-3}$.
    The flux model for each neutrino flavour is described by the luminosity (number of neutrinos per unit of time), the average neutrino energy and the spectral shape parameter as a function of time, energy and the emission direction with respect to the observer. For the fluxes used in this work, the chosen direction is the one along which the strongest effect of the SASI is predicted. This choice has no significant impact on the total number of events expected at the detector when compared to the flux averaged over the total solid angle. The corresponding particle fluences (time-integrated fluxes) are shown in Figure~\ref{fig:garching-fluence} for the three CCSN progenitors considered for the sensitivity estimation. The total flux of the non-electronic neutrino flavours, $\nu_{x} = \{ \nu_{\mu},\nu_{\tau},\overline{\nu}_{\mu},\overline{\nu}_{\tau} \}$, is expected to be equally divided across the four species.
    
    The reference distance to the source is taken as $\SI{10}{kpc}$. The spectrum is integrated over the time duration given by the limit of the simulation, different for each progenitor. 
    
    Flavour conversion inside the supernova can result in significant changes of the relative flavour composition of the flux, depending on the neutrino mass ordering. The net result of this effect depends on the CCSN energy spectrum. Given the KM3NeT sensitivity in this energy regime, the detector simulation shows that a full flavour conversion of the $\nuebar$ flux, expected in the case of inverted ordering, produces a variation in the number of detected events of about +20\% for the $\SI{11}{\solarmass}$ and $-$20\% for the $\SI{27}{\solarmass}$ progenitor. The net result emerges from the balance between the lower luminosity and the higher mean energy of the swapped flux. The case of normal ordering corresponds to an intermediate case between the non-oscillated flux and the full flavor conversion for $\nuebar$. This effect is from hereon ignored, and the non-oscillated fluxes are considered as benchmarks for the presented analyses.
    
\section{The KM3NeT detectors}
\label{s:detector}

 KM3NeT is a research infrastructure under construction in the Mediterranean Sea. It consists of two deep-sea Cherenkov neutrino detectors, ARCA and ORCA, located off the coast of Capo Passero (Italy) and Toulon (France), at depths of about 3500 and 2500 metres underwater, respectively~\cite{KM3NeT:2016-LoI}.

 The key component of the KM3NeT detectors is the \emph{Digital Optical Module} (DOM), consisting of a pressure-resistant glass sphere instrumented with 31 $\SI{80}{mm}$ diameter PMTs in a three-dimensional arrangement. The DOMs are connected in groups of eighteen to form vertical lines, called \emph{detection units} (DUs). The DUs are anchored to the sea bed and kept vertical by the buoyancy of the DOMs and by dedicated buoys located at the top. An array of 115 detection units forms a \emph{building block}. Each detection unit is connected to the seafloor infrastructure which provides the electrical power and optical data networks.

 The two KM3NeT detectors share the same technology, adopting different instrumentation densities optimised for their respective primary physics goals. ARCA aims at the discovery and observation of astrophysical neutrino sources at the TeV--PeV energy range. It is a $\si{\kilo\metre^3}$-scale detector of two building blocks, with a vertical spacing of $\SI{36}{\metre}$ between the DOMs and a $\SI{90}{m}$ horizontal distance between detection units, on average. The main goal of ORCA is the study of atmospheric neutrino oscillations in the \SIrange{1}{100}{GeV} energy range, primarily aimed to determine the neutrino mass ordering. The ORCA single building block instruments a \SIrange{6}{7}{Mton} volume of seawater, having on average a $\SI{9}{\metre}$ vertical spacing between the DOMs and a $\SI{20}{\metre}$ horizontal distance between the detection units.

 In the KM3NeT DOMs, the analog signals from the 31 PMTs are digitised by a custom front-end electronic board~\cite{KM3NeT:2019qol-CLB}. The hit times of Cherenkov photons generating a signal above a threshold equivalent to 0.3 photoelectrons are digitised with nanosecond resolution. Following an \emph{all data to shore} concept, no data reduction is applied offshore. All hits are transmitted to a computing farm onshore where they are filtered and processed with trigger algorithms. For this analysis, the main sources of background are radioactive decays in seawater (mainly $\kforty$), bioluminescence and atmospheric muons. A characterisation of these backgrounds is given in Refs. \cite{PPM-DOM,PPM-DU}. The average background hit rate is $\sim\SI{7}{kHz}$ per PMT, dominated by radioactive decays. Bioluminescence can cause localised and diffused increases of the hit rates, up to the MHz range. A \emph{high rate veto} logic is adopted in the front-end electronics to suppress the data acquisition of a PMT when its rate is detected above $\SI{20}{kHz}$ on a $\SI{100}{ms}$ timescale. While the overall detector efficiency is reduced in presence of high bioluminescence activity, the uptime is not impacted. The average fraction of PMTs in high rate veto is of a few per mille in ARCA and of a few per cent in ORCA. The corresponding reduction in efficiency is estimated to be in the same order of magnitude.
 
 The Cherenkov emissions from radioactive decays and atmospheric muons produce tightly time-correlated photons that are detected as nanosecond-scale coincidences between multiple PMTs of the same DOM. For the purpose of this analysis, a coincidence is defined by allowing a maximal time difference of $\SI{10}{ns}$ between the hit times. The number of PMTs hit in a coincidence is defined as the \emph{multiplicity}, $M$. While radioactive decays are detected locally, muons produce multiple causally connected coincidences on different DOMs along their paths. This criterion is used in the trigger to identify events caused by GeV--PeV neutrinos or atmospheric muons.

 PMT hit data are grouped in 100 ms time segments (\emph{timeslices}) that are processed onshore by parallel software data filters. Two types of data are available after the data filtering: \emph{triggered events} and \emph{timeslice data}. A triggered event is generated when at least one trigger algorithm has identified a cluster of causally connected coincidences matching a topology of interest. The latter can be cylindrical (\emph{track-like}) for muon tracks or spherical (\emph{shower-like}) for electromagnetic and hadronic cascades. The hits matched by the trigger as part of the physics signature are referred to as \emph{triggered hits}. A triggered event stores all the hit data (\emph{snapshot}) recorded by the detector in a time window that covers all the triggered hit times plus a designated margin. Timeslice data consist of all hit data for a selection of coincidences that simultaneously satisfy three configurable conditions: (i) a maximal time difference between the hit times, (ii) a minimum multiplicity, and (iii) a maximal opening angle between the corresponding PMT axes. Separate timeslice streams are generated with selections dedicated to different purposes and are subject to different storage policies. Two types are considered in this work. The sensitivity and energy estimations (Sections \ref{s:sensitivity} and \ref{s:energy}) are based on timeslices providing all the hits from coincidences with at least four hit PMTs within 10~ns and a 90 degree opening angle. The analyses of the neutrino signal time profile (Section~\ref{s:lc}) are instead based on a timeslice stream providing all coincidences with at least two hits on different PMTs within a 25~ns time window, without angular selection. A different (shorter) time window can be optionally adopted in the subsequent analysis of hit data.

\section{Detection of CCSN neutrinos in KM3NeT}
\label{s:ccsn-km3net}

The spacing between the optical modules in a KM3NeT detector allows the reconstruction only of sufficiently extended or bright events, above a threshold of few $\si{GeV}$s. The interaction of a neutrino below $\SI{100}{MeV}$ produces a charged lepton ($e^{+}$ or $e^{-}$) travelling up to a few tens of centimetres ($\sim\SI{0.5}{cm}$ per MeV of the incident neutrino energy~\cite{ICSN}). Since this distance is small compared to the typical separation between the KM3NeT DOMs, the corresponding Cherenkov signatures cannot be reconstructed as individual events. The detection of CCSN neutrinos relies on the observation of a population of coincidences in excess over the background expectation, taking into account all the DOMs in the detector. The multiplicity distribution of the detected coincidences is exploited to discriminate their origin on a statistical basis. The background rates are measured from the data acquired with the first KM3NeT detection units deployed in the sea. Given that the identifiable detection on multiple DOMs occurs for a negligible fraction of signal events, this kind of correlation is used instead to identify and subtract the contribution of atmospheric muons, exploiting the KM3NeT physics trigger algorithms. The efficiency of the atmospheric muon rejection is evaluated by applying the filter to simulated radioactivity and atmospheric muon events in a KM3NeT building block. The KM3NeT sensitivity to CCSN explosions is determined by comparing the background rates to the simulated signal of CCSN neutrinos on a single DOM.

\subsection{Simulation of CCSN neutrino interactions}
\label{ss:simulation}

 For the simulation of the CCSN neutrino signal in KM3NeT, the following interaction channels of low-energy neutrinos in water are considered:
 
 \begin{itemize}
 \item inverse beta decay (IBD) of electron anti-neutrinos on free protons ($\nuebar + p \rightarrow e^{+} + n$). It is the main detection process for water-based detectors~\cite{Scholberg:2012id}. In the case of KM3NeT, it accounts for $\sim88-93\%$ of the detection rate. This channel is favoured by its relatively large cross section and by the fact that the incident neutrino energy is efficiently transferred to the outgoing positron, enhancing the probability of detection;
 \item elastic scattering on electrons ($\nu + e^{-} \to \nu + e^{-}$), which is possible for all neutrino flavours and contributes at the $\sim3-5\%$ level;
 \item charged-current neutrino interactions with oxygen nuclei  ($\nu_e + \ce{ ^{16}O } \to e^{-} + \ce{ ^{16}\textrm{F} }$, $\nuebar + \ce{ ^{16}O } \to e^{+} + \ce{ ^{16}\textrm{N} } $). They contribute from 2\% up to 8\% to the detection rate, depending on the progenitor; \item neutral-current interaction with oxygen, inducing excited states resulting in de-excitation $\gamma$ photons, are neglected.
 \end{itemize}

 Neutrino interactions are generated using a custom software accounting for the energy-dependent cross sections and full event kinematics. Cross sections are taken from Ref.~\cite{Strumia:2003zx} for inverse beta decay, Ref.~\cite{Radel:1993sw} for elastic scattering and from~\cite{Kolbe:2002gk} for oxygen. The outgoing leptons produced in neutrino interactions are propagated in seawater with KM3Sim~\cite{Tsirigotis:2011zza}, a detailed simulation based on GEANT4~\cite{GEANT}. The lepton energy loss, the production of Cherenkov light as well as the photon propagation, absorption and scattering in seawater are taken into account. The angular acceptance, the wavelength-dependent quantum efficiency of the PMTs and the absorption in the DOM glass and optical gel are also considered. Finally, the detected photons are further processed through custom KM3NeT software, reproducing the analog PMT response, the readout electronics and the assembly of the raw data streams. The same data filtering and triggering algorithms used for real data are applied to the simulated raw data streams, producing an output format equivalent to the one of the KM3NeT data acquisition system.

 From simulated CCSN data, timeslices are processed to determine the number of detected coincidences (\emph{signal events}). The expected number of signal events as a function of the multiplicity is shown in Table~\ref{t:mult_ev} for the $\SI{11}{\solarmass}$, $\SI{27}{\solarmass}$ and $\SI{40}{\solarmass}$ CCSN progenitors. The number of events interacting through a process in a certain volume of water is given by the product of the neutrino flux and the interaction cross section, integrated over a time window $\delta t$:
 \begin{equation}
     N_{int, \kappa} = n_{\kappa} \int_E \int_{t=0}^{\delta t} \frac{d\Phi}{dE\,dt}(E,t) \, \sigma_{\kappa}(E) \, dE \, dt %
     \: ,
 \end{equation}
 where $\kappa \in \{ p, e^{-}, ^{16}\ce{O} \}$ represents the target, $n_{\kappa}$ is the number of targets, $\sigma_{\kappa}(E)$ the total interaction cross section for a given target and $d\Phi/(dE\,dt)$ the flux from Equation~\ref{eq:dphi-de}. The number of events is summed over all the interaction channels. Neutrino interactions are simulated in a spherical volume of $\SI{20}{\meter}$ radius centred on one optical module. The contribution to coincidences of interactions occurring beyond this radius is negligible. The detection efficiency is represented by the \emph{effective mass}, namely the water mass of a detector with unit efficiency. It corresponds to the ratio between the number of detected events and the number of interacting neutrinos per unit of mass, and is calculated as:
 \begin{equation}
    M_{\rm{eff}}(M) = \frac{N_{\rm{det}}(M)}{N_{\rm{int}}} \rho_{\rm{water}} V_{\rm{gen}} \: , 
 \end{equation}
 where $\rho_{\rm{water}}$ is the water density, $N_{\rm{det}} (M)$ is the number of detected events at multiplicity $M$ in the simulation, and $N_{\rm{int}}$ is the number of neutrinos interacting inside the generation volume, which has size $V_{\rm{gen}}$, for the simulated CCSN flux. The total generation volume corresponds to the volume of a sphere of 20~m radius multiplied by the number of KM3NeT optical modules.
 
 For one building block, this corresponds to a total generation mass of 69~Gton of water. For the three progenitors, Table~\ref{t:meff} provides the corresponding effective mass as a function of the multiplicity for one KM3NeT building block. The effective masses corresponding to the event selections used in the sensitivity estimation (Section~\ref{s:sensitivity}) and in the time profile study (Section~\ref{s:lc}) are of $\sim$ \SIrange{0.7}{1.5}{\kilo ton} (multiplicity 7--11) and $\sim$ \SIrange{40}{67}{\kilo ton} (all coincidences), respectively.

\begin{table*}
\begin{center}
\caption{Expected number of signal events as a function of the multiplicity for one KM3NeT building block (2070 DOMs) for the three different progenitors considered at 10~kpc. The errors on the expected values coming from the statistical uncertainty of the Monte Carlo simulation are reported. For each progenitor, the duration of the corresponding flux simulation is indicated in parentheses.}
\label{t:mult_ev}
\resizebox{0.77\textwidth}{!}{
\begin{tabular}{cccccc}
 \toprule
  \multirow{2}{*}{\textbf{Model}} & \multicolumn{5}{c}{\textbf{Multiplicity}} \\ \cmidrule(l){2-6} 
  & {\textbf{2}} & {\textbf{3}} & {\textbf{4}} & {\textbf{5}} & {\textbf{6}} \\
 \midrule
 $\SI{11}{\solarmass}$ ($\SI{340}{ms}$) & $ 1119 \pm 3 $ & $ 258 \pm 1 $ & $ 100.4 \pm 0.8 $ & $ 48.9 \pm 0.5 $ & $ 25.8 \pm 0.4 $ \\
 \midrule
 $\SI{27}{\solarmass}$ ($\SI{543}{ms}$) & $4806 \pm 9$ & $1120 \pm 5$ & $442 \pm 3$ & $218 \pm 2$ & $116.0 \pm 1.5$ \\
 \midrule
 $\SI{40}{\solarmass}$ ($\SI{572}{ms}$) & $15240 \pm 30$ & $3650 \pm  10$ & $1449 \pm 8$ & $723 \pm 6$ & $399 \pm 4$ \\

  \cmidrule(l){2-6} & {\textbf{7}} & {\textbf{8}}  & {\textbf{9}} & {\textbf{10}} & {\textbf{11}} \\
 \midrule
 $\SI{11}{\solarmass}$ ($\SI{340}{ms}$) & $ 13.3 \pm 0.3 $ & $ 7.2 \pm 0.2 $ & $ 3.4 \pm 0.1$  & $ 1.29 \pm 0.08$ & $0.50 \pm 0.05$ \\
 \midrule
 $\SI{27}{\solarmass}$ ($\SI{543}{ms}$) & $64 \pm 1$ & $35.2 \pm 0.8$ & $19.4 \pm 0.6$ & $8.0 \pm 0.4$ & $1.9 \pm 0.2$ \\
 \midrule
 $\SI{40}{\solarmass}$ ($\SI{572}{ms}$) & $226 \pm 3$ & $127 \pm 2$ & $69.5 \pm 1.8$ & $36.6 \pm 1.3$ & $15.0 \pm 0.8$ \\
 \bottomrule
 \end{tabular}
}
\end{center}
\end{table*}

\begin{table}
\begin{center}
\caption{Effective mass (in kton) as a function of the multiplicity for the $\SI{11}{\solarmass}$ ($\avg{E_{\nu}}=\SI{13.7}{MeV}$), $\SI{27}{\solarmass}$ ($\avg{E_{\nu}}=\SI{15.7}{MeV}$) and $\SI{40}{\solarmass}$ ($\avg{E_{\nu}}=\SI{18.2}{MeV}$) progenitors. A systematic uncertainty of the order of 10\% should be assumed on the values (see Section~\ref{s:systematics}).}

\label{t:meff}

\begin{tabular}{cccccc}
 \toprule
  \multirow{2}{*}{\textbf{Model}} & \multicolumn{5}{c}{\textbf{Multiplicity}} \\ \cmidrule(l){2-6} 
  & {\textbf{2}} & {\textbf{3}} & {\textbf{4}} & {\textbf{5}} & {\textbf{6}} \\
 \midrule
 $\SI{11}{\solarmass}$ & 28 & 6.5 & 2.5 & 1.2 & 0.65 \\ 
 \midrule
 $\SI{27}{\solarmass}$ & 37 & 8.6 & 3.4 & 1.7 & 0.91 \\
 \midrule
 $\SI{40}{\solarmass}$ & 47 & 11 & 4.5 & 2.2 & 1.2 \\

  \cmidrule(l){2-6} & {\textbf{7}} & {\textbf{8}}  & {\textbf{9}} & {\textbf{10}} & {\textbf{11}} \\
 \midrule
 $\SI{11}{\solarmass}$ & 0.34 & 0.18 & 0.094 & 0.034 & 0.015\\
 \midrule
 $\SI{27}{\solarmass}$ & 0.49 & 0.276 & 0.15 & 0.069 & 0.025\\
 \midrule
 $\SI{40}{\solarmass}$ & 0.70 & 0.40 & 0.21 & 0.11 & 0.052 \\
 \bottomrule
 \end{tabular}


\end{center}
\end{table}

\subsection{Optical background measurement}

 For a KM3NeT DOM, the background rate as a function of the multiplicity is characterised by the distribution presented in Ref.~\cite{KM3NeT:2019-MuonDepth}. Radioactive decays dominate at low multiplicities, with a rate of $\sim\SI{500}{Hz}$ at multiplicity 2, roughly decreasing by an order of magnitude for every step in multiplicity. At multiplicity 6, the contribution of atmospheric muons becomes relevant, dominating at 8 and above.
 
\begin{figure}
     \centering
     \includegraphics[width=\columnwidth]{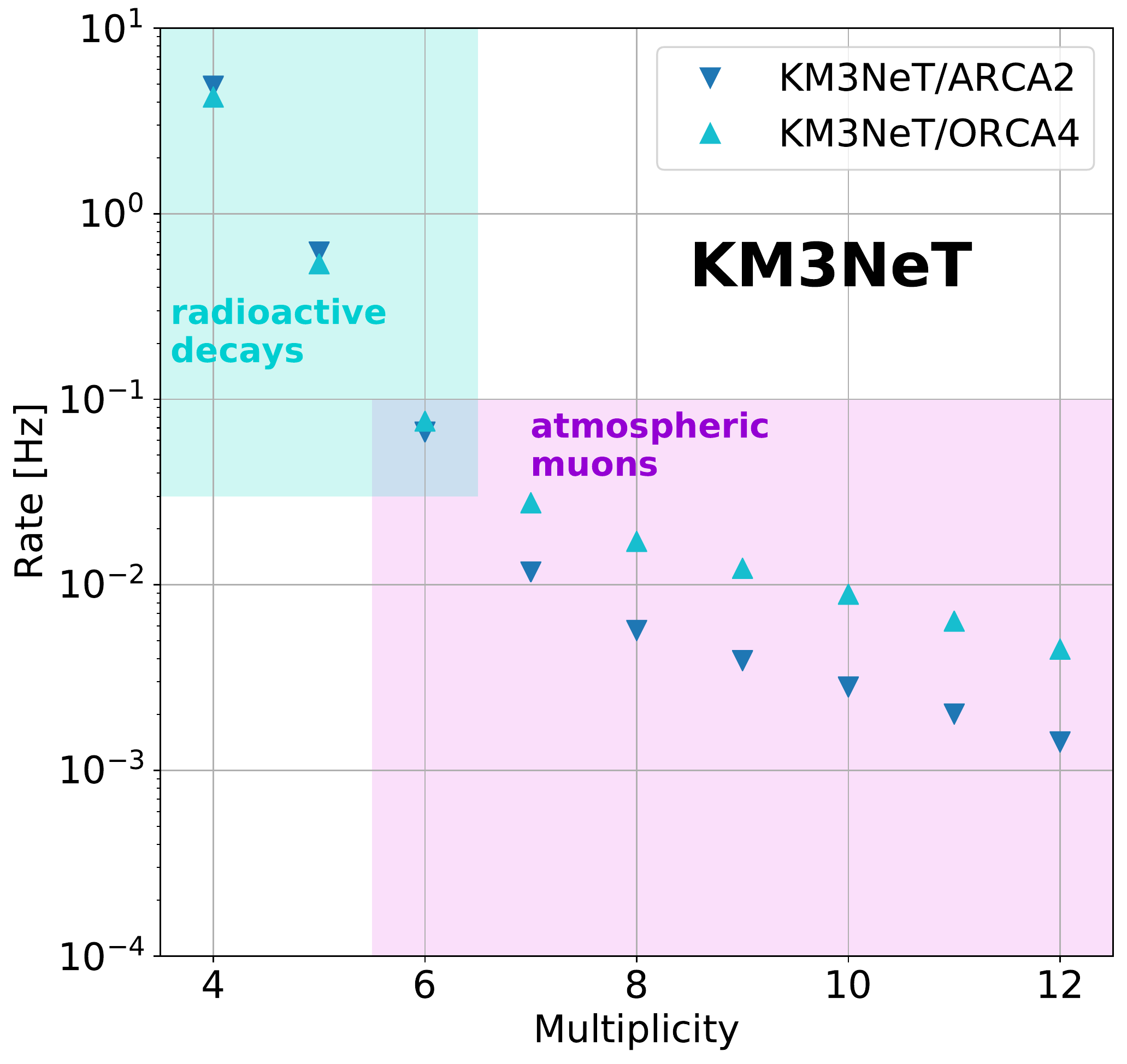}
     \caption{Average DOM coincidence rate as a function of the multiplicity measured with the KM3NeT ARCA2 and ORCA4 detectors. Statistical errors are included and are smaller than the markers. The main contribution to the coincidence rate is indicated with the shaded areas. The multiplicity range shown here covers the most relevant region for CCSN neutrino detection.}
     \label{fig:optical-background}
 \end{figure}

The average optical background rates for a KM3NeT DOM are shown in Figure~\ref{fig:optical-background}. The rates have been measured from the data of the first two deployed lines of ARCA (ARCA2) and the first four deployed lines of ORCA (ORCA4). The selected data taking periods are from December 23, 2016 to March 2, 2017 for ARCA2 and from September 30, 2019 to November 4, 2019 for ORCA4. In the considered periods, the detectors showed stable photon detection efficiencies. The reference background rates have been estimated selecting timeslices for which at least $99\%$ of the PMTs were active (i.e.~not suppressed by the high rate veto logic).
 
\subsection{Background filtering}
 The background filtering strategy has two aims: reducing the contribution of the optical noise and suppressing the detection of multiple coincidences corresponding to the same background event, e.g., in the case of atmospheric muon interactions (affecting different DOMs) and PMT afterpulses (affecting a single DOM).
 
 Bioluminescence emission is a single-photon process that can only contribute to random coincidences. For a 10~ns coincidence time window, these are negligible above multiplicity two. Bioluminescence can, however, impact the overall efficiency of the detector, as the signals from the PMTs which are above the high rate veto threshold are suppressed. The effect of this veto condition is discussed in Section~\ref{s:systematics}. For radioactive decays, since the energy of the emitted electron is an order of magnitude lower with respect to CCSN neutrinos, a cut on the minimum multiplicity is the most robust and effective reduction strategy.

 As introduced above, the background from atmospheric muons can be reduced by exploiting the fact that muon tracks typically produce correlated coincidences on multiple DOMs. The KM3NeT trigger algorithms are designed to identify a minimum number of causally connected hits within extended cylindrical sections or localised spherical sections of the instrumented volume~\cite{KM3NeT:2016-LoI}. In this analysis, for each triggered event produced by the data filter, a \emph{veto} is applied on the set of DOMs detecting at least one triggered hit. The veto lasts for the total duration of the event, as defined by the time range of the triggered hit times. Typical values for this interval are \SIrange{1}{3}{\micro\second}. The remaining coincidences are analysed on a DOM-by-DOM basis. If one or more coincidences occur on a DOM within $\SI{1}{\micro\second}$, only the coincidence with the highest multiplicity is kept. This selection results in an effective reduction of the background rates and in the suppression of spurious coincidences. The effectiveness of the approach is verified on data taken with the ARCA2 and ORCA4 detectors operated in the sea. The $\si{\micro\second}$-scale average duration of a muon veto multiplied by a muon trigger rate of $\sim\SI{100}{\hertz}$ per building block results dead time fraction below $\num{E-3}$, which is negligible.
 
\begin{figure}
    \centering
    \includegraphics[width=\columnwidth]{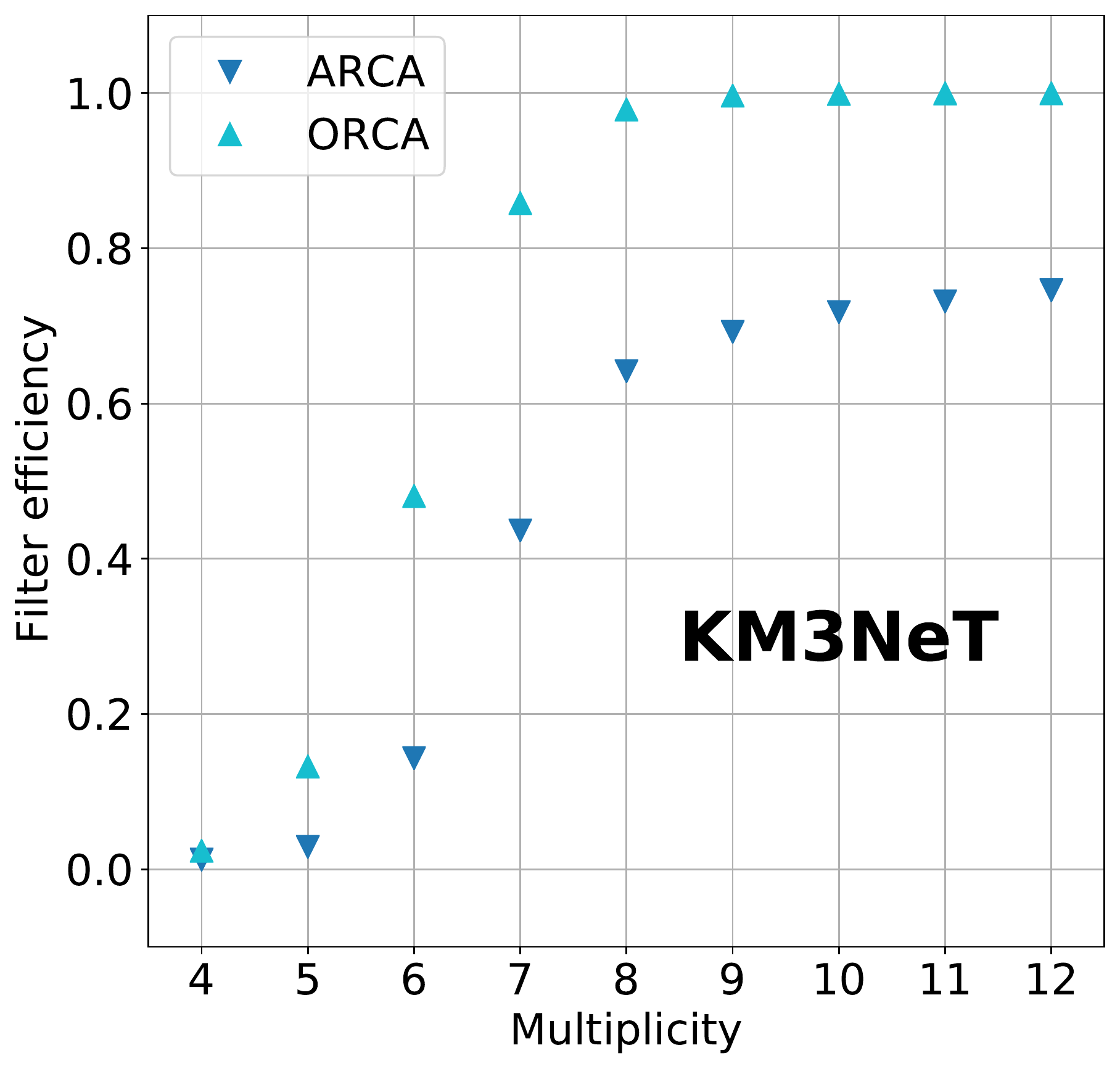}
    \caption{Fraction of coincidences rejected by the background filter, estimated with simulations of the ARCA and ORCA building blocks.}
    \label{fig:filter-efficiency}
\end{figure}

 The efficiency of the background rejection is evaluated for one ARCA and one ORCA building block with Monte Carlo simulations. The simulation chain is based on the atmospheric muon event generator MUPAGE~\cite{Becherini:2005sr} and a Cherenkov light simulator implemented in the custom KM3NeT software. The generation of the simulated hit data follows the procedure outlined in Ref.~\cite{KM3NeT:2019-MuonDepth}. The fraction of coincidences rejected by the filter as a function of the multiplicity is shown in Figure~\ref{fig:filter-efficiency}. ARCA reaches a $65\%$ rejection efficiency at multiplicity eight and above. In the same range, the denser geometry of the ORCA detector allows for the identification and suppression of more than $95\%$ of the background. The difference is due to the fact that lower-energy muons are not triggered in ARCA as efficiently as in ORCA. The impact of the filtering strategy on the signal is negligible, since the low-energy CCSN neutrino interactions do not significantly contribute to the trigger rate.
 
\section{Detection sensitivity}
\label{s:sensitivity}

 The sensitivity of KM3NeT to a CCSN neutrino burst is evaluated considering the number of signal and background events in a $\SI{500}{\milli\second}$ time window after the onset of the core-collapse. This time window is chosen as it corresponds to the typical duration of the accretion phase $\overline{\nu}_{e}$ burst, as shown in~\cite{Totani:1997vj,Mirizzi:2015eza}. Here, the assumption is that the arrival time of the burst at the detector is known from an independent observation tied to the time of the core-collapse, such as a neutrino signal detected by another detector or a gravitational-wave burst. The length of the time window is chosen to cover, on average, the majority of the neutrino emission occurring in the accretion phase.

 In order to be compared with the signal simulation, the measured background rates (Figure~\ref{fig:optical-background}) are corrected for the average photon detection efficiency of the PMTs of each detector. Then, the efficiency of the muon background rejection estimated in the ARCA and ORCA building block simulations (Figure~\ref{fig:filter-efficiency}) is applied to the corrected rates to obtain the event rate of the background as a function of the multiplicity.

 For the $27$ and $\SI{40}{\solarmass}$ progenitors, the interaction rate in the $\SI{500}{ms}$ time window starting at the core bounce is used to compute the expected number of signal events at the detector. In the case of the $\SI{11}{\solarmass}$ progenitor, the rate is extrapolated beyond the end of the simulated interval considering a constant value between 340 and $\SI{500}{ms}$, as it is seen in the case of the $\SI{27}{\solarmass}$ progenitor and 1D simulations~\cite{Mirizzi:2015eza}, covering the time evolution of the CCSN for a longer duration.
 
\begin{figure}[ht]
    \centering
    \includegraphics[width=\columnwidth]{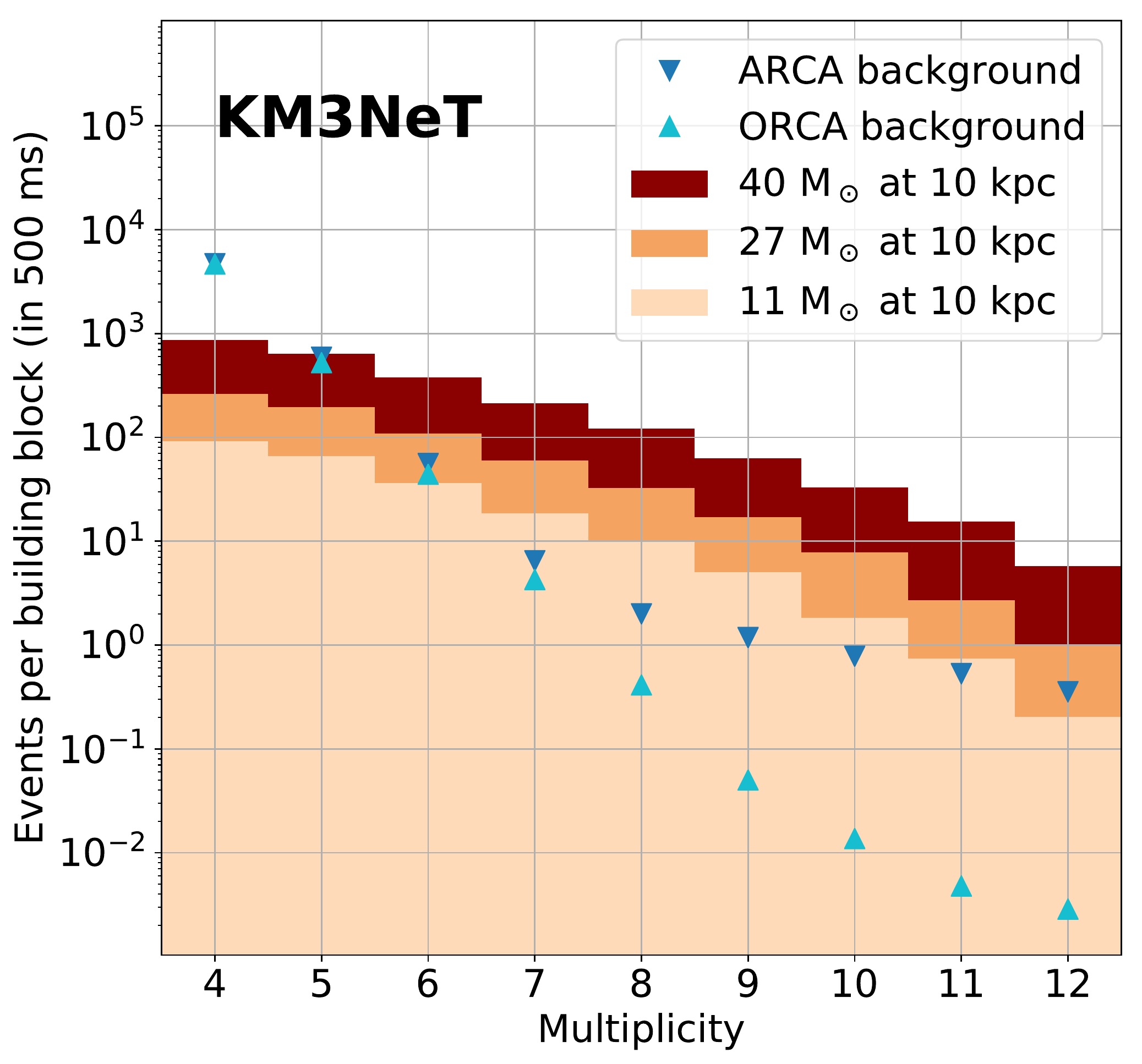}
    \caption{Expected number of events in a KM3NeT building block as a function of the multiplicity. The background is shown with markers in light blue for ORCA and dark blue for ARCA. The signal is represented with coloured bars in orange shades for the different models: light for $\SI{11}{\solarmass}$, intermediate for $\SI{27}{\solarmass}$, dark for $\SI{40}{\solarmass}$.}
    \label{fig:snr}
\end{figure}

 In Figure~\ref{fig:snr}, the number of expected events in a $\SI{500}{ms}$ time interval for a single KM3NeT building block of 2070 DOMs is reported. The estimated backgrounds in ARCA and ORCA are compared with the simulated signal for the $\SI{11}{\solarmass}$, $\SI{27}{\solarmass}$ and $\SI{40}{\solarmass}$ CCSN progenitors. Due to the difference in the muon filter performance, the background rate at higher multiplicity is lower in ORCA than in ARCA. From hereon, the computations account for the respective size of the complete KM3NeT detectors: two building blocks for ARCA and one for ORCA.

 The sensitivity of a Poisson counting experiment to a given signal hypothesis can be defined as the expected median significance of its observation. In the large sample limit, the sensitivity, expressed in terms of Gaussian standard deviations, can be approximated by the formula~\cite{Cowan:2010js}:
 \begin{equation}
 \label{eq:sensitivity}
    Z = \sqrt{2 \left((n_s + n_b)\,\ln \left( 1 + \frac{n_s}{n_b} \right)  - n_s \right)} ,
 \end{equation}
 where $n_s$ and $n_b$ are the expectation values for the number of signal and background events, respectively.
 
\begin{figure*}[!ht]
     \centering
     \includegraphics[width=\textwidth]{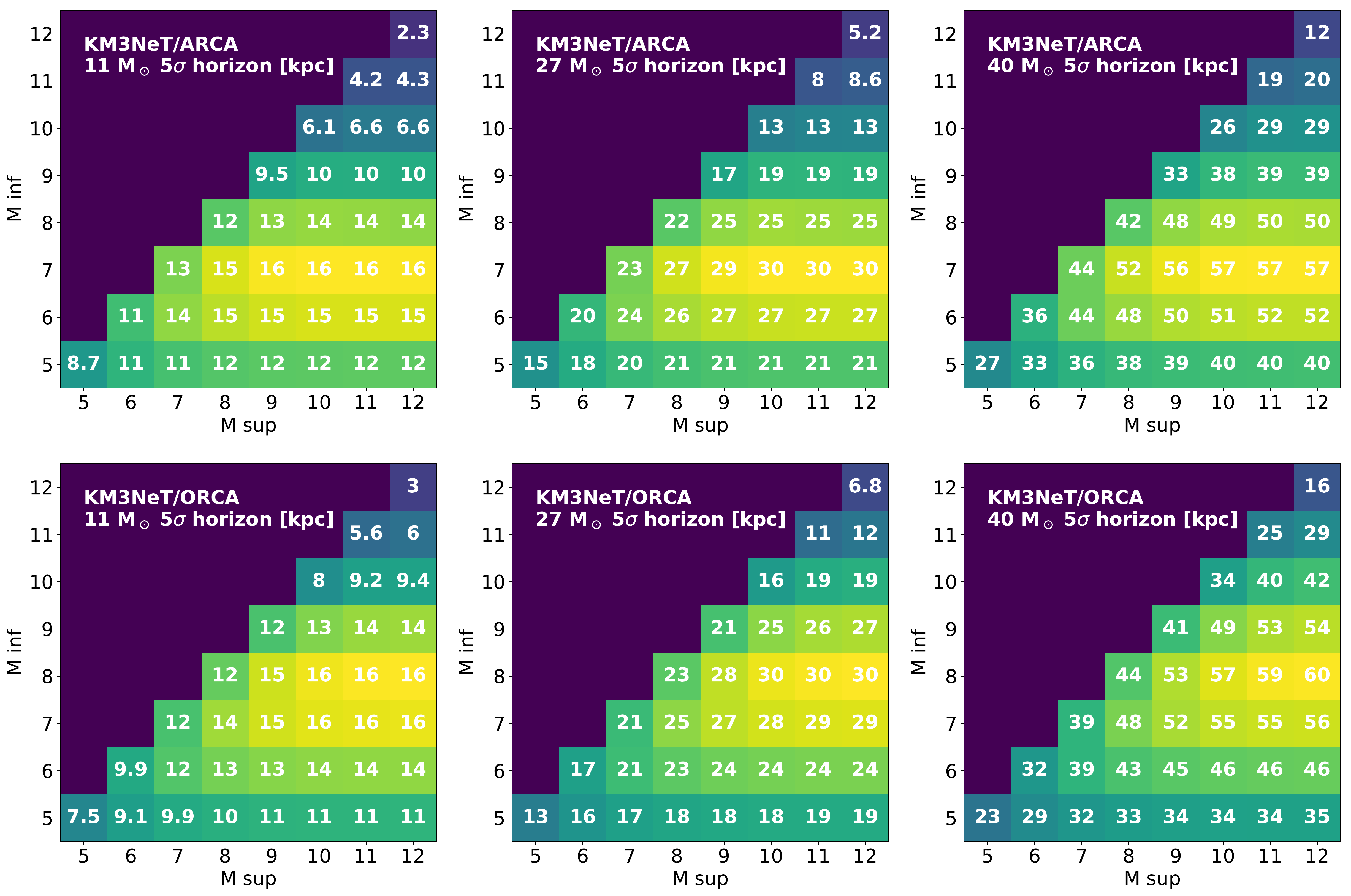}
     \label{fig:msel}
     \caption{$5\,\sigma$ discovery horizons (box numbers) estimated for the $\SI{11}{\solarmass}$ (left), $\SI{27}{\solarmass}$ (middle) and $\SI{40}{\solarmass}$ (right) CCSN progenitors as a function of the minimum ($M_{\rm inf}$) and maximum ($M_{\rm sup}$) multiplicity. Top for the ARCA detector, bottom for the ORCA detector.}
 \end{figure*}

 Considering the number of signal events as a function of the distance $n_s (d) = n_s(d_0) (d_0/d)^2 $, with $d_0 =$~$\SI{10}{kpc}$, the $5\,\sigma$ discovery distance is evaluated as a function of the minimum and maximum multiplicities. Figure~\ref{fig:msel} reports the results for the three progenitors and the two detectors, taken individually. The optimal sensitivity is achieved across the 7--10(12) multiplicity ranges for ARCA and in the (7)8--10(12) multiplicity ranges for ORCA, the parentheses indicating that the same sensitivity is reached for both cuts.

 The final multiplicity range is chosen taking into account qualitative considerations. If the minimum multiplicity cut is too high, the available statistics for the background is significantly reduced, preventing an exhaustive evaluation of the method reliability with currently available data. In addition, if the mean energy of the CCSN neutrinos is lower than the worst case considered here, a higher number of events would be expected at lower multiplicities. A cut on the maximum multiplicity is, on the other hand, adopted to exclude a region where the signal contribution is negligible (see Table~\ref{t:mult_ev}) and a statistically reliable evaluation of the background stability is not possible. On the basis of these considerations, the 7--11 multiplicity range is adopted for both detectors.

 Considering $Z$ in Equation~\eqref{s:sensitivity} as a function of the distance, $Z(d)$, the KM3NeT combined sensitivity is obtained by a weighted linear combination of the ORCA and ARCA sensitivities:
 \begin{equation}
    \label{eq:combination}
    Z_{\textrm{KM3NeT}}(d) = \frac{\sum_{\alpha \in \{\textrm{ARCA}, \textrm{ORCA}\}} w_\alpha Z_\alpha (d)}{\sqrt{\sum_{\alpha \in \{\textrm{ARCA}, \textrm{ORCA}\}} w_\alpha^2}},
 \end{equation}
 where the weight, $w$, is defined as the sensitivity at a reference distance of $\SI{10}{kpc}$, $w_\alpha = Z_\alpha(d = \SI{10}{kpc})$.

\begin{figure*}[!ht]
    \centering
    \includegraphics[width=0.75\textwidth]{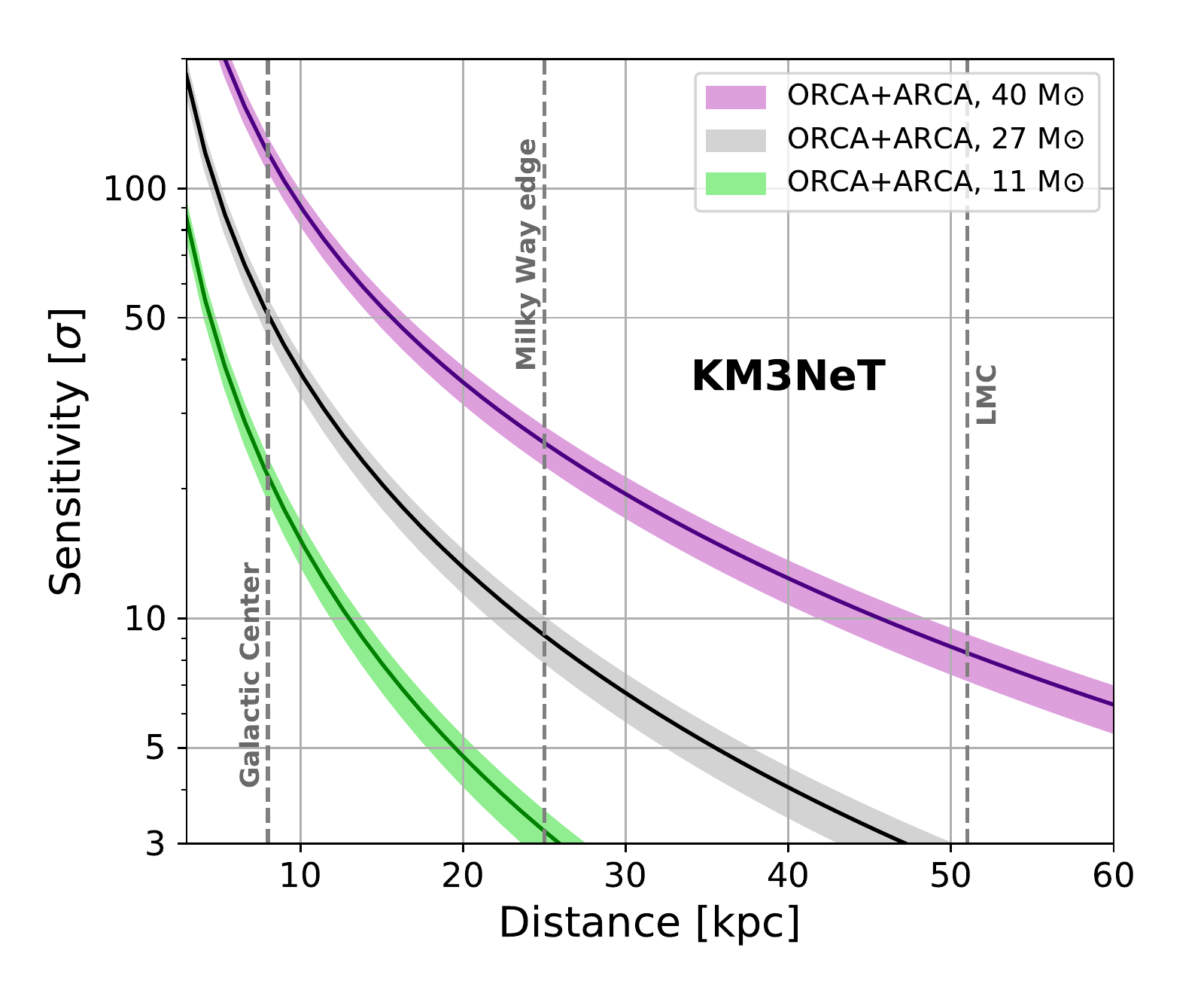}
    \caption{KM3NeT detection sensitivity as a function of the distance to the CCSN for the three progenitors considered: $\SI{11}{\solarmass}$ (green), $\SI{27}{\solarmass}$ (black) and $\SI{40}{\solarmass}$ (purple). The error bars include the systematic uncertainties summarised in Section~\ref{s:systematics}.}
    \label{fig:sensitivity}
\end{figure*}

The number of signal and background events at 10~kpc after the background filter for the chosen multiplicity range, together with the detection sensitivities, are given in Table~\ref{t:events} for each progenitor and for the two KM3NeT detectors. In Figure~\ref{fig:sensitivity}, the sensitivity for the combination of the ORCA and ARCA detectors is reported as a function of the distance to the source for the three considered progenitors.

Taking into account the expected distribution of CCSNe as a function of the distance to the Earth \cite{Adams:2013ana}, in the most conservative scenario considered in this paper ($\SI{11}{\solarmass}$), more than 95\% of the Galactic core-collapse supernovae can be observed by the KM3NeT detectors. KM3NeT will thus contribute to the observation of the next Galactic explosion. The sensitivity to the black-hole forming case ($\SI{40}{\solarmass}$) extends beyond the Large Magellanic Cloud. By comparison, the most sensitive detectors currently in operation, such as IceCube~\cite{ICSN,ICSN_ICRC2019} and Super-Kamiokande~\cite{SKSN_proba}, can typically detect a CCSN up to the Large and Small Magellanic Clouds.

\begin{table}[!ht]
\begin{center}

\caption{Expectation values for the number of background and signal events after the background rejection in the chosen 7--11 multiplicity range for ARCA and ORCA. The signal is given for a CCSN at a reference distance of 10~kpc. The corresponding sensitivity for each detector and progenitor is provided.}
\label{t:events}

\resizebox{\columnwidth}{!}{%
\begin{tabular}{ccccccc}\toprule
\multirow{2}{*}{\bf Progenitor} & \multicolumn{3}{c}{\bf ARCA} & \multicolumn{3}{c}{\bf ORCA} \\ \cmidrule(l){2-7} 
& $N_b$ & $N_s$ & $\sigma_{\SI{10}{kpc}}$ & $N_b$ & $N_s$ & $\sigma_{\SI{10}{kpc}}$
 \\ \midrule
$\SI{11}{\solarmass}$ & 22.1 & 72.2 & 11 & 4.9 & 36.1  & 10 \\
 \midrule
$\SI{27}{\solarmass}$ & 22.1 & 240 & 29 & 4.9  & 120  & 24 \\ \midrule
$\SI{40}{\solarmass}$ & 22.1 & 895 & 71 & 4.9 & 447  & 57 \\
\bottomrule
\end{tabular}
}
\end{center}
\end{table}

\begin{figure*}
    \centering
    \includegraphics[width=0.75\textwidth]{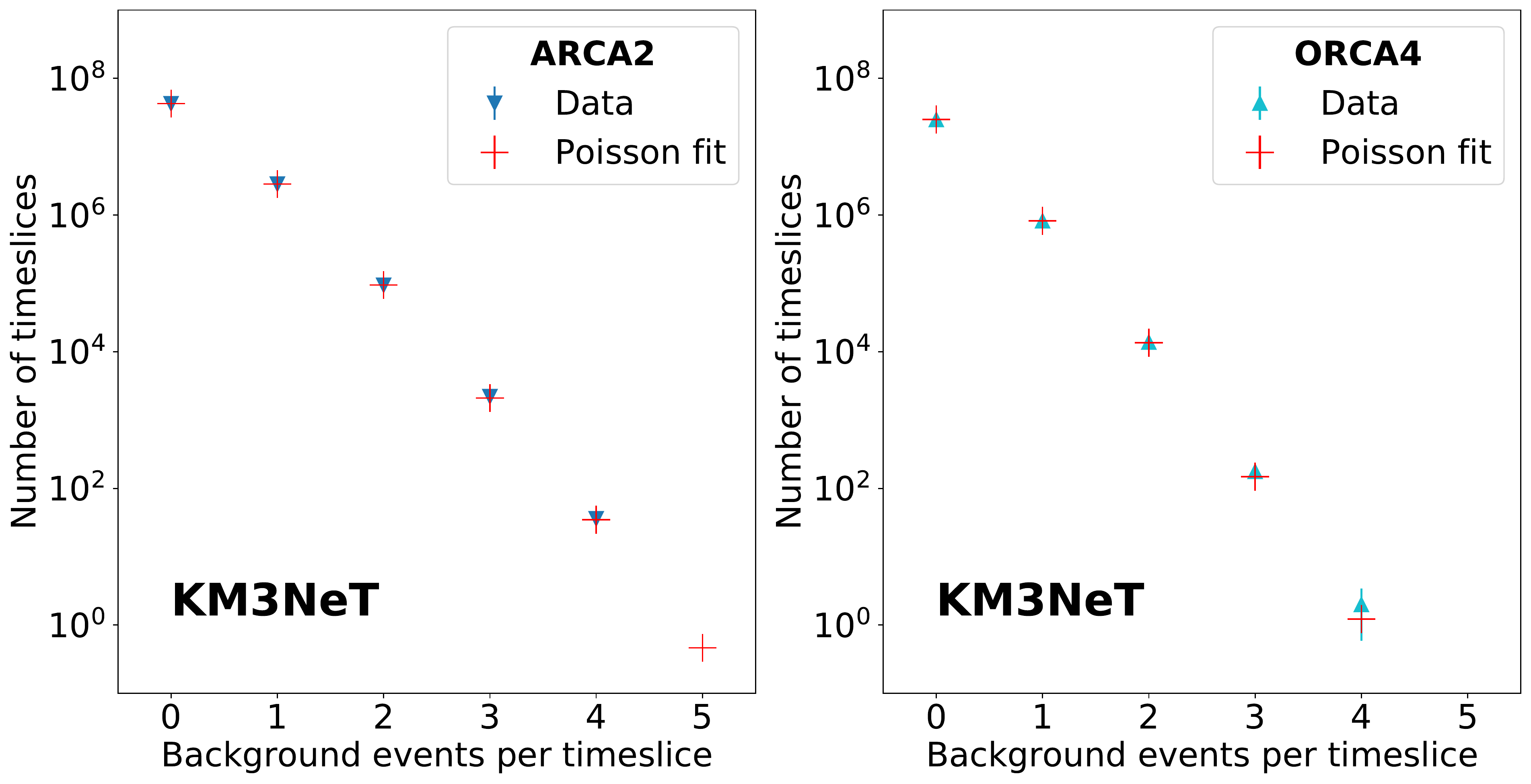}
    \caption{Number of $\SI{100}{ms}$ timeslices as a function of the number of the background events in the timeslice for the considered ARCA2 (left) and ORCA4 (right) data taking periods. Statistical errors are included. A Poisson distribution with expectation value equal to the mean value of the data is overlayed with red markers.}
    \label{fig:background-sampling}
\end{figure*}

 The estimation of the sensitivity assumes that the number of signal and background events on the time scale of the CCSN search ($\SI{500}{ms}$) is distributed according to the Poisson statistics. The number of background events after the filter is evaluated in the chosen 7--11 multiplicity range for all the $\SI{100}{ms}$ timeslices in the considered data taking periods of ARCA2 and ORCA4. In Figure~\ref{fig:background-sampling}, the number of timeslices as a function of the number of background events detected in the timeslice is shown. For this, timeslices with a fraction of active PMTs (i.e. not suppressed by the high rate veto logic) above 85\% are considered. The Poisson distribution with expectation value equal to the mean of the data sample is also drawn. The data are found to be compatible with the Poisson statistics and do not show outliers.

\section{Systematic uncertainties}
\label{s:systematics}
 Assuming the flux models introduced in Section~\ref{s:flux-models}, all the relevant systematic uncertainties affecting the results have been evaluated. This includes the uncertainties on the determination of the neutrino interaction rate and the corresponding detection efficiency, as well as the uncertainties affecting the evaluation of the background expectation. These uncertainties will be assessed here as a percentage indicating the relative variation in the expected number of signal and background events.
 
 The interaction rate, and therefore the expected number of signal events, depends on the cross sections of the different processes. For inverse beta decay and electron elastic scattering, these are known with sub-percent accuracy. The higher uncertainty for interactions with oxygen nuclei can be neglected due to the small contribution of this channel to the signal.
 
 Water absorption length can impact the number of detected photons per signal event. The overall uncertainty on this property is assumed to be 10\% in KM3NeT studies~\cite{KM3NeT:2016-LoI,Riccobene:2006qy}. The effect of this variation has been evaluated through simulations to 3\% for multiplicity two and 1\% for multiplicity seven and above. This behaviour is expected, as detected events producing higher multiplicity have their vertices closer to a DOM.
 
 The absolute PMT efficiency has an impact on the number of detected photons per event. From calibration studies, an uncertainty of $\pm 5\%$ has been determined \cite{KM3NeT:2019-MuonDepth}. In the 7--11 multiplicity range, the corresponding effect on the coincidence rates induced by CCSN neutrinos has been evaluated to be within $\pm 10\%$.

 A finite generation volume, consisting of a sphere of 20~m radius centred on the DOM, is used to simulate the signal events. The contribution of neutrinos interacting outside this volume is estimated to be lower than $1\%$ for an extension of $\SI{5}{m}$ of the sphere radius.
 
 The uncertainties described above do not affect the estimation of the background rates, as their measurement is based on data.

 Due to the high rate veto logic, the number of active PMTs changes as a function of time. This variation is translated into a reduction of the overall efficiency of the detector, and therefore of the expected number of background events. From the sole knowledge of the number of active PMTs, this effect can be estimated with an error of $\pm 3\%$. This uncertainty is applied to the total number of signal and background events.

 The efficiency of the background filter is evaluated with Monte Carlo simulations of one ARCA and one ORCA building blocks. The comparison of data and Monte Carlo for the ARCA2 detector shows that the filter is $15\%$ less efficient in data for the considered multiplicity range. This deviation is accounted for as a systematic uncertainty on the background rate.

 A summary of the results for the different systematic uncertainties studied is shown in Table~\ref{t:systematics}. 
 \begin{table}[!ht]
 \begin{center}
 \caption{Systematic uncertainties evaluated in this work. The first column represents the variable under study while the second one indicates the percentage of variation evaluated. The third column indicates the corresponding uncertainties for the 7--11 multiplicity range. The percentages indicate the variation of the signal (S) and background (B) expectations associated to each systematic uncertainty.}
 \label{t:systematics}
 \resizebox{\columnwidth}{!}{%
 \begin{tabular}{ccc}\toprule   
 {\bf Variable} & {\bf Variation} & {\bf Systematic uncertainty} \\ \midrule
 PMT efficiency & $\pm 5\%$ & (S) $\pm$ 10\%  \\ \midrule
 Active PMTs & $\pm$3\% & (S, B) $\pm$3\%  \\ \midrule
 Finite generation radius & $+ \SI{5}{m}$ & (S) $<$1\%  \\ \midrule
 Absorption length & $\pm$10\% & (S) $\pm$ 1\%  \\ \midrule
 IBD/ES cross sections & $<1\%$ & (S) $<$1\%  \\ \midrule
 Filter efficiency & & (B) $+ 15\%$ \\
\bottomrule
\end{tabular}}
\end{center}
\end{table}

\section{Estimation of the neutrino spectrum parameters}
\label{s:energy}

The sensitivity to the neutrino energy spectrum is estimated using a CCSN flux described by Equation~\ref{eq:dphi-de}, considering perfect flavour equipartition and no time variation of the spectrum. 
The simulated data from ARCA and ORCA are combined in a $\SI{500}{ms}$ search window. The neutrino spectrum is characterised by three parameters: $\avg{E}$, $\alpha$, and the signal scale, $\Lambda$. The signal scale depends on the total energy released and the distance to the source. It is defined with respect to the benchmark values, $L_{ \overline{\nu}_{e} ,0} = \SI{4E52}{erg}$ and $d_{0}=\SI{10}{\kilo\parsec}$, as:
\begin{equation}
    \Lambda = \frac{L_{ \overline{\nu}_{e} }}{L_{ \overline{\nu}_{e} ,0}} \, \left( \frac{d_{0}}{d} \right)^{2}\: .
\end{equation}

\begin{figure}[!ht]
    \centering
    \includegraphics[width=\columnwidth]{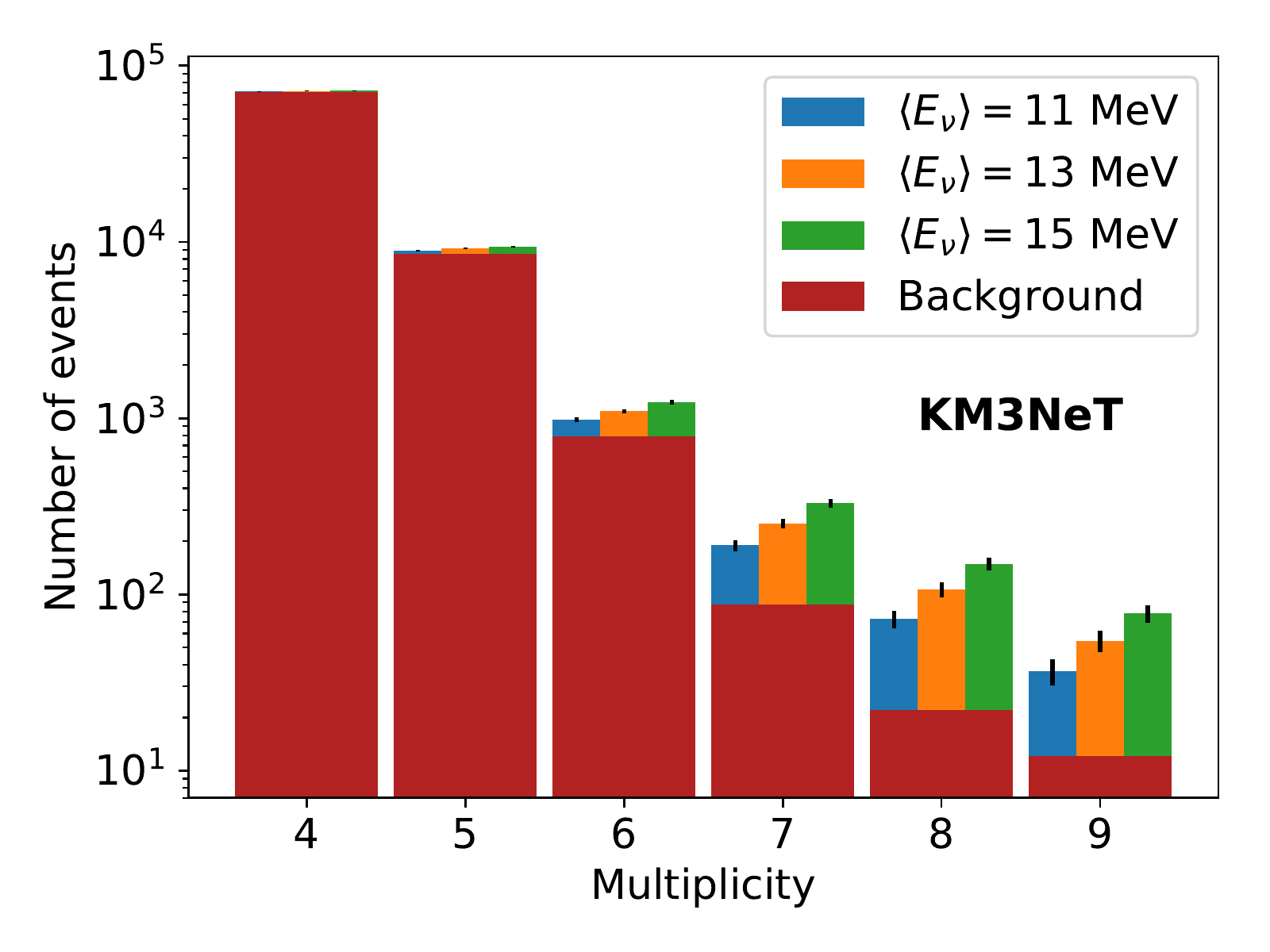}
    \caption{Expected multiplicity distributions of detected events in ARCA and ORCA detectors for CCSN $\bar\nu_e$ spectra having mean energies $\avg{E_{\nu}}$ = 11, 13 and $\SI{15}{MeV}$, $\alpha = 3$ and $\Lambda = 1$. The number of events due to the background of the detector is drawn in red.}
    \label{fig:mrates}
\end{figure}

The analysis strategy exploits the fact that the observed multiplicity distribution depends on the flux spectral features. In particular, neutrinos with higher energies tend to produce more photons and be detected as coincidences with higher multiplicities. This is illustrated in Figure~\ref{fig:mrates}. The highest multiplicity considered here is lowered to 9 compared to Section~\ref{s:sensitivity}, since for  higher multiplicities the event statistics becomes very low, and does not provide a stable contribution.

The ability to constrain the parameters describing the neutrino energy spectrum is evaluated using a chi-square method. Pseudo-experiments with fixed true values for the three parameters are performed to construct the probability density function, $h_\chi$, of the $\Delta\chi^2$ defined as:
\begin{equation}
\Delta\chi^2=\chi^{2}(\avg{E}_\mathrm{true},\alpha_\mathrm{true}, \Lambda_\mathrm{true})-\chi^{2}(\hat{\avg{E}},\hat\alpha, \hat{\Lambda}) \:, 
\end{equation}
where $\hat{\avg{E}}$, $\hat{\alpha}$ and $\hat{\Lambda}$ are the parameter values that minimise the $\chi^{2}$.
This distribution is used to define the $\Delta\chi^2_\mathrm{crit}$ value that corresponds to the 90\% confidence level (CL) as follows:
\begin{equation}
    \int_0^{\Delta\chi^2_\mathrm{crit}} h_\chi(\Delta\chi^2) \, d(\Delta\chi^2) \le 0.9 \; .
\end{equation}

\begin{figure}[!ht]
    \centering
    \includegraphics[width=\columnwidth]{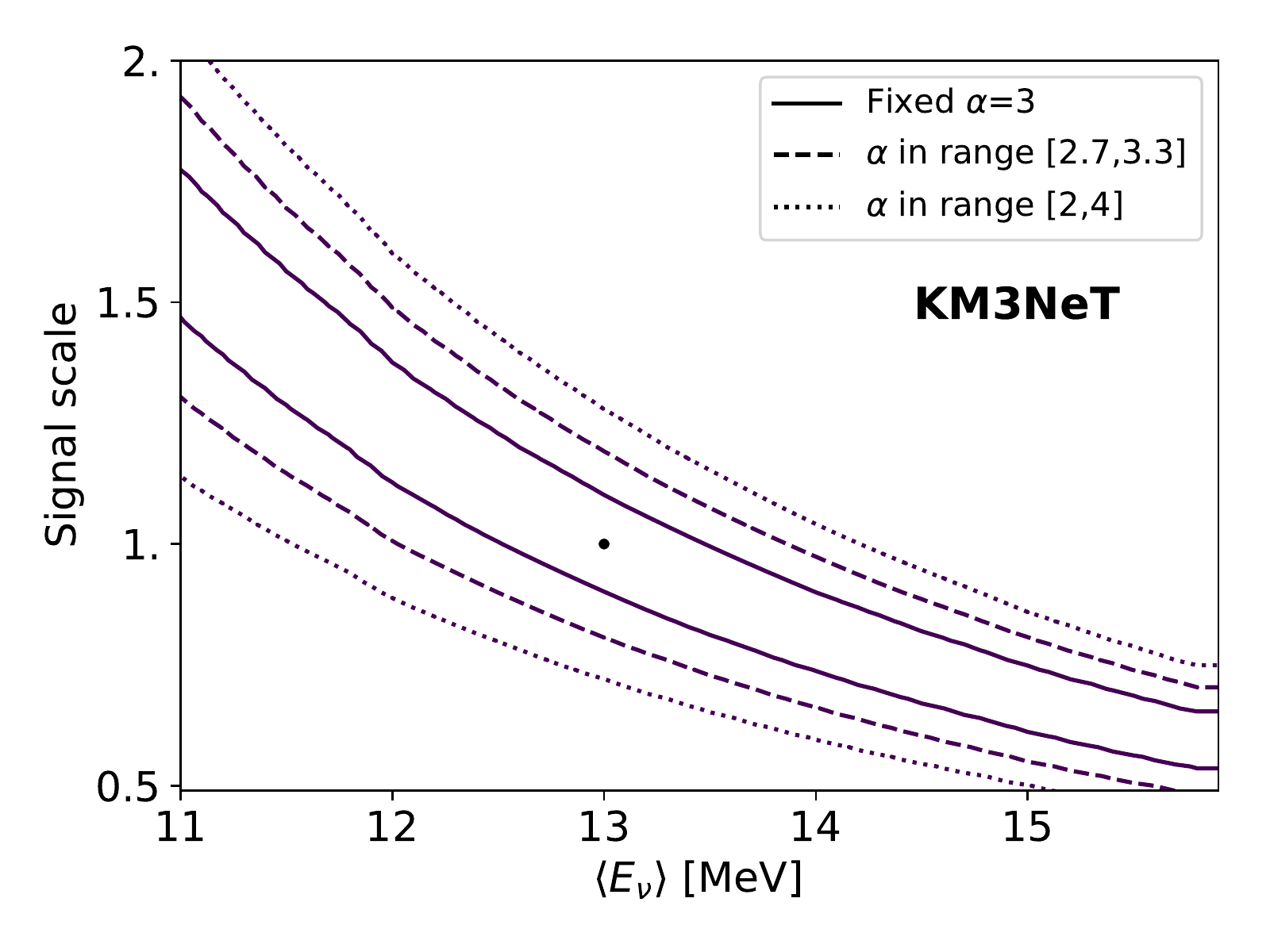}
    \caption{Contours at 90\% confidence level in the signal scale, $\Lambda$, and $\avg{E_{\nu}}$ parameter space for the assumed combined ARCA and ORCA data sets in a 500~ms time window. Three hypotheses are considered for the spectral shape parameter: $\alpha$ free in the fit in the full range of expected values 2--4 (dotted line), $\alpha$ in the range $\alpha_{\rm true}$ $\pm$ 10\% (dashed line), and $\alpha$ fixed to the true value (solid line). The dot indicates the true values.}
    \label{fig:encontours}
\end{figure}

Confidence level contours are defined as the subset of the parameter space $(\avg{E},\alpha, \Lambda)$ of the $\nuebar$ spectrum for which $\chi^{2}(\avg{E},\alpha, \Lambda)-\chi^{2}(\hat{\avg{E}},\hat\alpha, \hat{\Lambda}) \le \Delta\chi^2_\mathrm{crit}$. The \emph{Asimov} data set~\cite{Cowan:2010js} is used to evaluate the 90\% CL contours for a signal hypothesis having true values $\alpha_{\rm true} = 3$, $\avg{E}_{\rm true} = \SI{13}{MeV}$ and $\Lambda_{\rm true} = 1$. Three different assumptions on the range of the $\alpha$ parameter are considered: $\alpha$ is a free parameter in the range of 2--4, $\alpha$ is free in a constrained range given by $\alpha_{\rm true} \pm 10\%$, and $\alpha$ is known (fixed). The results are shown in Figure~\ref{fig:encontours}.

The results are the following: the mean neutrino energy estimation has a 90\% CL range of about $\pm$ 0.5~MeV ($\sim4\%$) when $\alpha$ and $\Lambda$ are fixed, i.e. known \emph{a priori}, and stays below $\pm$ 1.5~MeV if these two parameters are known within a 10\% variation around the true value. The sensitivity to the spectral parameters is lost if $\Lambda$ and $\alpha$ are left free.

Alternatively to the confidence areas obtained from the true data sets, the distributions of the fitted values from simulated pseudo-experiments are built to estimate the precision in fitting the neutrino spectral parameters. With this method, the uncertainty on the parameters estimation is given by the width of the distribution (RMS) of the difference between the fitted and the true values. The analysis is performed for three different hypotheses on the shape parameter and the signal scale: first assuming $\alpha$ and $\Lambda$ are precisely known, second limiting the range to the true value $\pm$ 10\% for both parameters, and third assuming $\Lambda$ being free in the range $[0,2.5]$, and considering a physically allowed range for $\alpha$ being $[2,4]$. For the first case, the mean neutrino energy resolution is 0.25~MeV ($\sim2\%$). With the limited range (case 2), the mean energy uncertainty remains of about 0.4~MeV ($\sim3\%$), while most of the fitted values for $\alpha$ and $\Lambda$ reach the fitting range edges, making the estimate of these two parameters unreliable. In the third case, when $\Lambda$ and $\alpha$ are fitted in the larger range of possible values, the sensitivity to the spectral parameters is completely lost.

For comparison, the results reported by other sensitive Cherenkov experiments are summarised. As in the case of KM3NeT, the IceCube detector is not able to do an event-by-event analysis. However, coincidences between different optical modules can be used to estimate the mean neutrino energy~\cite{ICSNenergy,Bruijn:2013ibl,Demiroers:2011am}. Assuming the other parameters fixed, the mean energy uncertainty is of $\sim30\%$, which is an order of magnitude worse than the KM3NeT performance presented here. With the coming IceCube-upgrade, coincidences detected by multi-PMT optical modules can be exploited, improving the energy resolution up to about 5\%~\cite{Aartsen:2020fgd}. The capabilities of Super-Kamiokande and Hyper-Kamiokande to resolve the CCSN spectrum in an event-by-event reconstruction using the Cherenkov light patterns have been explored in Ref.~\cite{alphaHKSK}. A fit to the three spectral parameters is performed. For Super-Kamiokande (Hyper-Kamiokande) the results obtained on the $\nuebar$ spectrum show an accuracy on the mean neutrino energy of 6\% (2\%), no resolution (7\%) on $\alpha$, and  10\% (4\%) on the total energy. In comparison, KM3NeT has no resolution if all three parameters are left free.

\section{Time profile of the neutrino burst}
\label{s:lc}
 
 In case of a high-significance detection, the large event statistics collected by KM3NeT can be exploited in a detailed analysis of the time profile of the neutrino burst with a potential millisecond time resolution. The reconstruction of the time profile of the signal can enable the triangulation of the source, either by the independent estimation of the burst arrival times at different detectors \cite{Burrow:triang} or by directly comparing the experimental neutrino light curves \cite{Coleiro:2020vyj}. The analysis of the time profile can also be a powerful tool for model discrimination, especially when considering black-hole forming scenarios that exhibit an abrupt interruption of the neutrino emission. In addition to its own analysis capabilities, KM3NeT will be able to promptly share the neutrino light curve data with multi-messenger networks such as SNEWS2.0~\cite{SNEWS2.0}.
 
 In this section, two analyses are presented: the determination of the arrival time of the burst and the detection of oscillations in the neutrino light curve, as produced by the standing accretion shock instability.
 
 The coincidence selection used in Section~\ref{s:sensitivity}, from hereon referred to as \emph{CCSN selection}, aims to maximise the detection sensitivity providing a high purity sample. In this section, all coincidences recorded in the detector with at least two different hit PMTs are considered to investigate time-dependent properties. The corresponding expected number of signal events and the effective mass are reported in Tables~\ref{t:mult_ev} and \ref{t:meff}, respectively. The typical background rate is $\sim\SI{500}{Hz}$ per DOM from genuine coincidences induced by radioactivity. To this, the random combinatorial contribution of uncorrelated single hits is added. In this analysis, a reduced $\SI{5}{\nano\second}$ time window is adopted, decreasing the random contribution to half its value with respect to the $\SI{10}{\nano\second}$ case. For the simplified assumption of a $\sim\SI{7}{\kilo\hertz}$ single-hit rate per PMT, the rate of random coincidences is $\sim\SI{225}{Hz}$ per DOM. The corresponding loss of signal rate from the reduced coincidence window is estimated to be of the order of 3\%.

 Time-dependent variations of the background rate can be induced by bioluminescence, introducing auto-correlation in the background time profile. To realistically account for this, the background samples are simulated starting from measured time-dependent rates. The data from the ARCA and ORCA DUs are analysed to estimate the total coincidence rate as a function of time, with millisecond resolution. To emulate the background that would be observed in a full detector, multiple sequences of the measured rates are stacked. For example, $115$ sequences of $\SI{500}{ms}$, measured with one detection unit, are added to obtain a $\SI{500}{ms}$ sample for a 115-lines building block.
 
 The detected neutrino light curve is simulated by generating a background sample and adding a Poisson realisation of the signal. The latter is based on the time dependent signal expectation evaluated through a complete simulation (see Section~\ref{ss:simulation}). For both analyses, the events in the obtained light curve are grouped in time bins of $\SI{1}{\milli\second}$. For the SASI analysis, pure background samples are also generated.

\subsection{Arrival time of the CCSN neutrino signal}
\label{s:t0}

 The determination of the arrival time of the neutrino burst is of interest both for the astronomical and the astrophysical aspects of the CCSNe study. The combination of the arrival times at different detectors around the Earth can be used to localise the source by triangulation. Not only neutrinos can act as an early warning for optical follow-ups, but may also reveal optically obscured supernovae occurring in dense regions of the Galaxy, like the Galactic Centre itself. A precise knowledge of the arrival time of the signal can also help reducing the search time window for a gravitational wave counterpart, that would be expected shortly after the core bounce \cite{MM_SN}. From the astrophysical perspective, the relative start time of the electron anti-neutrino signal with respect to the $\nu_{e}$ burst is tied to the flavour conversion processes in the star, that in turn depend on the neutrino mass ordering. The combination of accurate timing information from detectors sensitive to different neutrino flavours could help in the reconstruction of the time profile of the neutrino signal, potentially revealing the intrinsic properties of neutrinos or the core-collapse mechanism.

 Due to the distance between the two KM3NeT sites, the ORCA and ARCA detected neutrino light curves will have a relative time offset of up to 3~ms, dependent on the source localisation. As the latter is not known \emph{a priori}, the measurement of the time of arrival is here evaluated for the ARCA detector alone, that has the best expected performance being twice the effective mass of ORCA. The $\SI{11}{\solarmass}$ and $\SI{20}{\solarmass}$ CCSN progenitors are considered as a conservative and optimistic case, respectively.
 
 As proposed in Ref.~\cite{Hansen}, the arrival time of the burst, $T_0$, can be measured in a large volume neutrino detector by performing an exponential fit of the signal leading edge. Before the fit is applied, the time range of the fit and the starting point of the $T_0$ parameter need to be determined from the experimental data. For this purpose, the time distribution of the events in the CCSN selection is exploited in a first step.
 
 After subtracting the background expectation, the time profile is scanned with a $\SI{20}{\milli\second}$ moving sum in steps of $\SI{1}{\milli\second}$. The lower edge of the first $\SI{20}{\milli\second}$ time interval containing a signal excess of $2.5\,\sigma$ above the background expectation is taken as a first estimator, $T_{0}^{M}$, of the time of arrival of the burst. The uncertainty on $T_0^{M}$ is evaluated using pseudo-experiments, with the true arrival time randomised between zero and the 20~ms bin width. In the case of ARCA, the background rate of the CCSN selection is $\SI{200}{\hertz}$, translating into an expectation of 4 background events per $\SI{20}{\milli\second}$. $T_{0}^{M}$ is then defined as the lower edge of the first interval containing at least 11 events.

 The $T_{0}^{M}$ estimator is biased, i.e. it exhibits a time shift, $T_{\mathrm{shift}}$, with respect to the true value. This comes from the fact that the number of signal events in the first time bins expected from the CCSN selection is small and not distinguishable from the background fluctuations. The value of the shift depends on the signal scale, i.e. on the total number of detected neutrinos for a given progenitor, normalised to the squared distance to the source. In order to have a method independent on the signal scale, $T_{\mathrm{shift}}$ is first evaluated for the benchmark case of a $\SI{20}{\solarmass}$ progenitor at 5~kpc, having a reference value of $T_\mathrm{shift} \simeq \SI{25}{\milli\second}$. Considering the CCSN selection, the shift is rescaled with the ratio between the total number of events expected from the simulated benchmark model, $I_{\nu, 0}$, and the total number of observed signal events, $I_{\nu}$. The starting value of the $T_0$ parameter in the fit, $T_{\mathrm{start}}$, is then assigned as:
\begin{equation}
    \label{eq:tstart}
    T_{\mathrm{start}} = T_{0}^{M} - T_{\mathrm{shift}}\,I_{\nu,0}\, I_{\nu}^{-1} \: .
\end{equation}

 The corresponding time interval for the light curve fit is taken as: $[ T_{0}^{M} - \SI{150}{ms}, T_{0}^{M} + I_{\nu,0}\,I_{\nu}^{-1} \, \SI{50}{ms} ] \:$. The lower limit of the range is chosen to include a background region large enough to ensure the stability of the fit, reducing the impact of fluctuations. The upper limit is, on the other hand, restricted to avoid going beyond the accretion peak (see Figure~\ref{fig:LC1}), including signal features that could bias the fit.
 
 Before the fit of the light curve, the background expectation value is subtracted. As a consequence, the mean value of the event rate before the signal onset is zero. 
 A moving-average filter is applied to reduce the background fluctuations. The width of the averaging window, $w_b$, is adjusted to optimise the time resolution by using the same scaling factor introduced for Equation~\eqref{eq:tstart}: 
 \begin{equation}
    w_b = w_b^{0} \, I_{\nu, 0}\,I_{\nu}^{-1} \: .
 \end{equation}
 Starting from a value of $w_b^{0} = \SI{20}{ms}$, the tested values go from 20~ms to 60~ms for distances between 5~kpc and 9~kpc, when considering the $\SI{20}{\solarmass}$ progenitor.
 
 The $T_0$ parameter is estimated by fitting the detected neutrino light curve in the selected time interval with the function:
 \begin{equation}
 \label{eq:fit}
    R(t) = \Theta(t - T_0) \, R_0 \left(1 - e^{-\frac{t - T_0}{\tau}} \right)
 \end{equation}
 where $\Theta$ is the step function, $\tau$ is the time constant of the light curve rise, and $R_0$ the event rate at the end of the fitting interval.
 
\begin{figure}[!ht]
    \centering
    \includegraphics[width=\columnwidth]{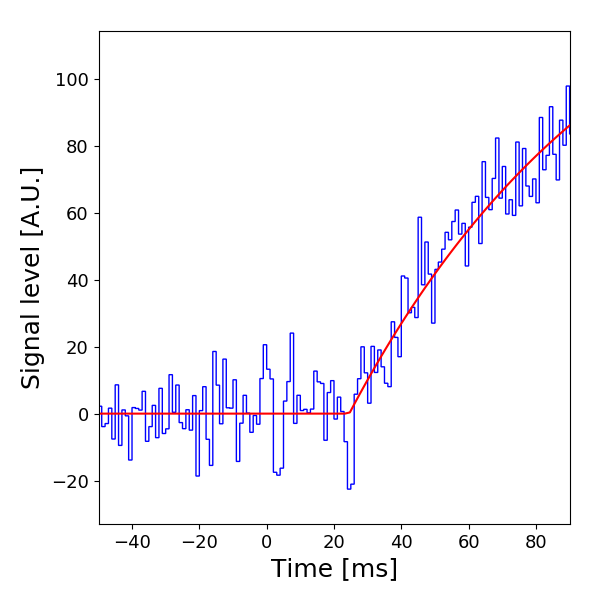}
    \caption{Time profile of the signal (light curve) in ARCA (two building blocks) using all coincidences for a $\SI{20}{\solarmass}$ CCSN progenitor at 5~kpc. The \emph{signal level} is obtained from the simulated experimental light curve, after subtracting the background and applying a moving-average filter with a 23~ms time window. The curve is fitted with the function in Equation~\ref{eq:fit}.}
    \label{fig:L1fit}
 \end{figure}

 The function is fitted to the detected neutrino light curve using a $\chi^2$ minimisation algorithm. An example of the fit applied to a light curve simulated for the ARCA detector is shown in Figure~\ref{fig:L1fit}. The time uncertainty is estimated through pseudo-experiments as the root mean square of the error.
 
 The main systematic uncertainties for the considered coincidence sample have also been evaluated and accounted for in the analysis. They are presented in the form of percentages indicating the relative variation in the expected number events. For the signal, the 5\% uncertainty on the photon detection efficiency of the PMTs translates to a 10\% variation. Correspondingly, the 10\% uncertainty on the absorption length has a 3\% effect. The impact of the bioluminescence conditions on the ARCA background estimate produces a variation on the rate at most of 3\%. The PMT efficiency uncertainty results in a 10\% change in the expected number of background events.

\begin{figure}[!ht]
    \includegraphics[width=\columnwidth]{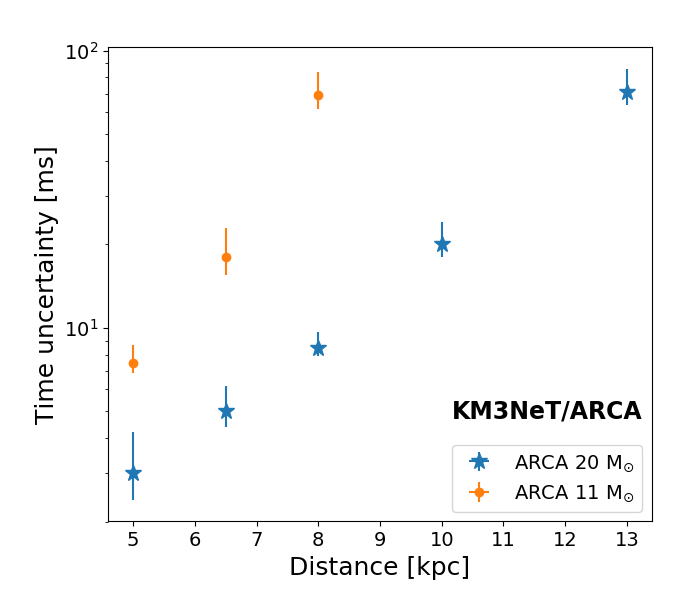}
    \caption{Uncertainty on the arrival time, $T_0$, estimated with the ARCA detector as a function of the distance to the source, assuming the $\SI{20}{\solarmass}$ and $\SI{11}{\solarmass}$ CCSN progenitors. The error bars include the most relevant systematic uncertainties.}
    \label{fig:t0result}
\end{figure}

 Figure~\ref{fig:t0result} provides the ARCA time uncertainty as a function of the distance, including the evaluated systematic uncertainties. An average time resolution of $\sim\SI{8}{ms}$ is achieved at the Galactic Centre ($\SI{8}{kpc}$) for the $\SI{20}{\solarmass}$ progenitor, improving to $\sim\SI{3}{ms}$ for an equivalent source at 5~kpc. At 13~kpc, the uncertainty degrades to $\sim\SI{70}{ms}$, with the fit failing $\sim25\%$ of the times. At 14~kpc, the estimation becomes unreliable as the fraction of failed fits reaches $\sim80\%$. For the $\SI{11}{\solarmass}$ progenitor, a resolution of $\sim\SI{7.5}{ms}$ is obtained at 5~kpc, degrading to about $\SI{70}{ms}$ at 8~kpc, with 35\% of failed fits. The fraction of failed fits increases to about 85\% at 9~kpc.

 For comparison, the IceCube detector can achieve a time uncertainty of $3-\SI{4}{ms}$ for a CCSN at 10~kpc ($\SI{13}{\solarmass}$ progenitor)~\cite{Cross:2019jpb}.

\subsection{Detection of the \emph{standing accretion shock instability} (SASI)}
\label{s:sasi}

As introduced in Section~\ref{s:flux-models}, the SASI phenomenon predicted by 3D simulations produces fast time variations in the neutrino light curve with a characteristic oscillation frequency.

\begin{figure*}[!ht]
    \centering
    \includegraphics[width=0.34\textwidth]{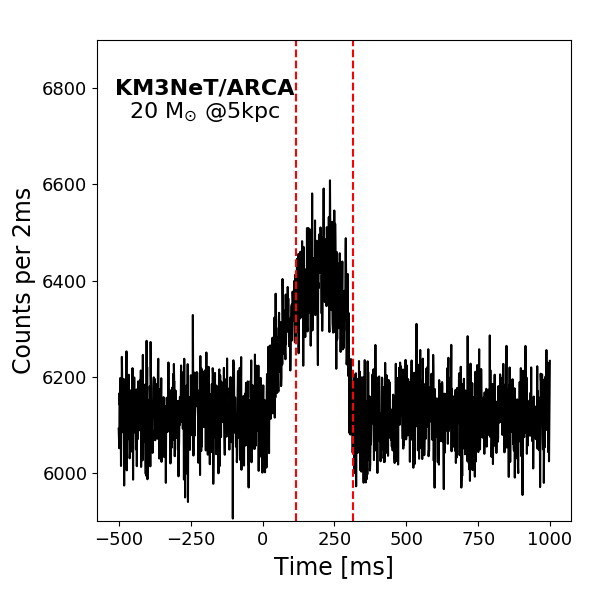}%
    \includegraphics[width=0.34\textwidth]{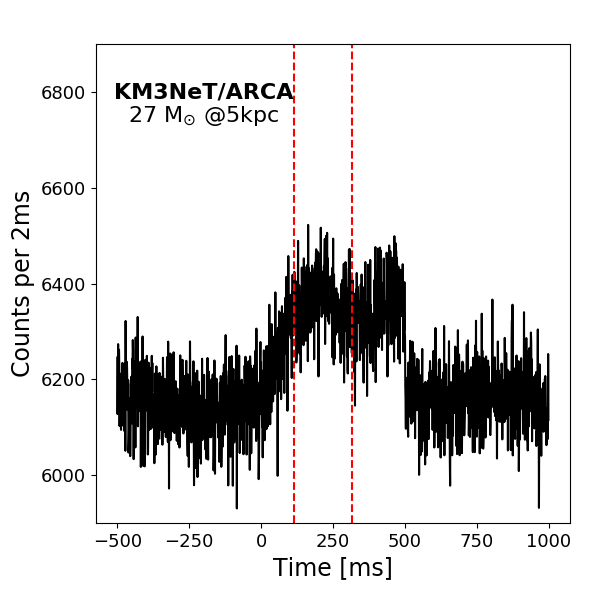}%
    \includegraphics[width=0.34\textwidth]{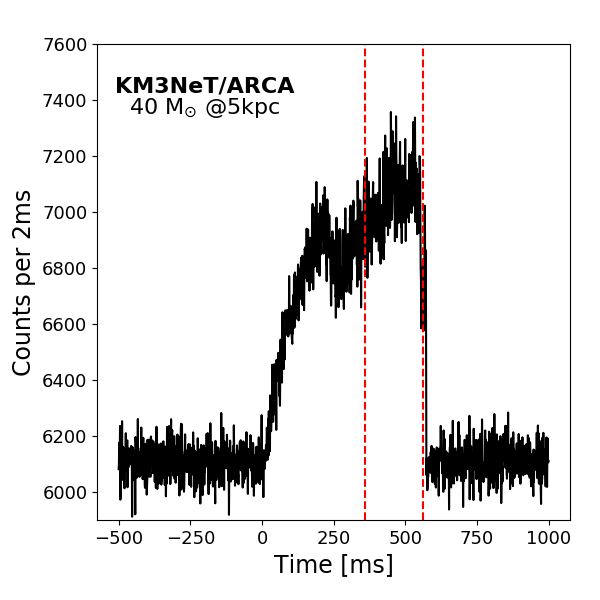}%
    \caption{Pseudo-experiments of the detected neutrino light curves in the full ARCA detector, considering a source at 5~kpc, and the three CCSN progenitors: the $\SI{20}{\solarmass}$ (left), $\SI{27}{\solarmass}$ (center), and $\SI{40}{\solarmass}$ (right). The dashed red lines indicate the interval to which the Fourier transform is applied.}
    \label{fig:LC1}
\end{figure*}

In this study, the $\SI{20}{\solarmass}$ and $\SI{27}{\solarmass}$ CCSN progenitors are taken into account. As the former shows an enhanced SASI activity with respect to the latter, the two progenitors can be considered as an optimistic and a more conservative CCSN scenario for this study, respectively. The $\SI{40}{\solarmass}$ is considered for the case of a failed CCSN leading to a black hole formation. Examples of the detected neutrino light curves obtained with pseudo-experiments are given in Figure~\ref{fig:LC1}. The light curve bin has been optimised to maximise the sensitivity to the SASI oscillation as of 2~ms.

A spectral analysis of the detected neutrino light curve is performed using a fast Fourier transform (FFT) algorithm. The procedure follows the approach used in Ref.~\cite{Tamborra:2013laa-SASI,Lund2010}. From the model prediction, a time interval of $[-\SI{150}{ms}, +\SI{50}{ms}]$ centred on the peak of the light curve is analysed. Given the length of the FFT window, $\tau = \SI{200}{ms}$, the corresponding spacing of the discrete Fourier frequencies is $\delta f = \tau^{-1} = \SI{5}{Hz}$.
To suppress boundary effects, a Hann windowing function is applied to the selected time interval. 

\begin{figure}
    \centering
    \includegraphics[width=0.9\columnwidth]{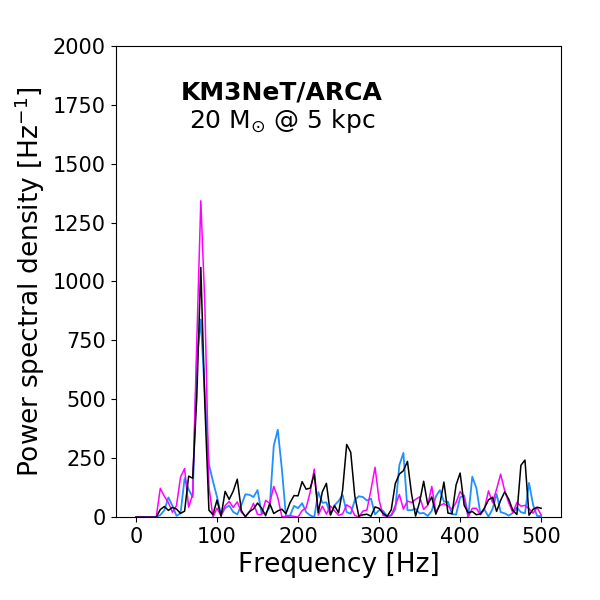}
    \caption{Power spectral densities for three simulated ARCA observations of the SASI from the $\SI{20}{\solarmass}$ progenitor CCSN signal at 5~kpc. The $\SI{80}{Hz}$ peak corresponds to the SASI frequency.}
    \label{fig:PS}
\end{figure}

In the following, an example of the analysis procedure is given for the ARCA detector, while the results will be evaluated for the combination of ORCA and ARCA. The corresponding power spectral densities (PSD) for three simulated observations of the $\SI{20}{\solarmass}$ progenitor are shown in Figure~\ref{fig:PS}.

From the power spectrum, two different strategies are adopted to estimate the probability of detecting the SASI oscillation. The first one (Method 1) is a model-independent search method based on the detection of a significant peak in the power spectral density (PSD$_{max}$). This approach is proposed to deal with the uncertainty on the expected SASI frequency for different progenitor models. The second model-dependent analysis (Method 2), is the search for a significant energy excess around a designated frequency, assumed \emph{a priori} according to the model prediction. A symmetric window of $\pm \SI{20}{Hz}$ centred on the assumed frequency is considered. The predicted SASI frequency is $\sim\SI{80}{Hz}$ for the $\SI{20}{\solarmass}$ and $\SI{27}{\solarmass}$ progenitors, and $\sim\SI{140}{Hz}$ for the $\SI{40}{\solarmass}$ CCSN. In the latter case, the SASI oscillations last for a longer period covering both the first and second peak of the light curve, while for the $\SI{20}{\solarmass}$ and $\SI{27}{\solarmass}$ they are only present around the first peak. In this analysis, regardless of the progenitor model, the search for the SASI oscillation uses a single time window centred on the neutrino light curve maximum, when the phenomenon has its peak intensity.

\begin{table*}[!b]
\caption{Sensitivity results to SASI oscillations obtained combining ORCA and ARCA for the three different stellar progenitors considered.}
\label{t:sasi}
\begin{center}
\begin{tabular}{ccccc}\toprule
{\bf Progenitor } & {\bf d [kpc] } & {\bf Method 1} & {\bf Method 2} & {\bf Galactic coverage} \\ \midrule 
27 M$_{\odot}$ & 3 &  $2.8 \pm 0.7\,\sigma$ &  $4.1 \pm 0.9\,\sigma$ & 3\% \\ \midrule
20 M$_{\odot}$ & 5 &  $3.2 \pm 0.7\,\sigma$ &  $4.5 \pm 0.9\,\sigma$ & 10\% \\ \midrule
40 M$_{\odot}$ & 8 &  $3.8 \pm 0.7\,\sigma$ &  $>$ $5\,\sigma$ & 35\% \\ \bottomrule
  \end{tabular}
\end{center}
\end{table*}

\begin{figure*}
    \centering
    \includegraphics[width=0.45\textwidth]{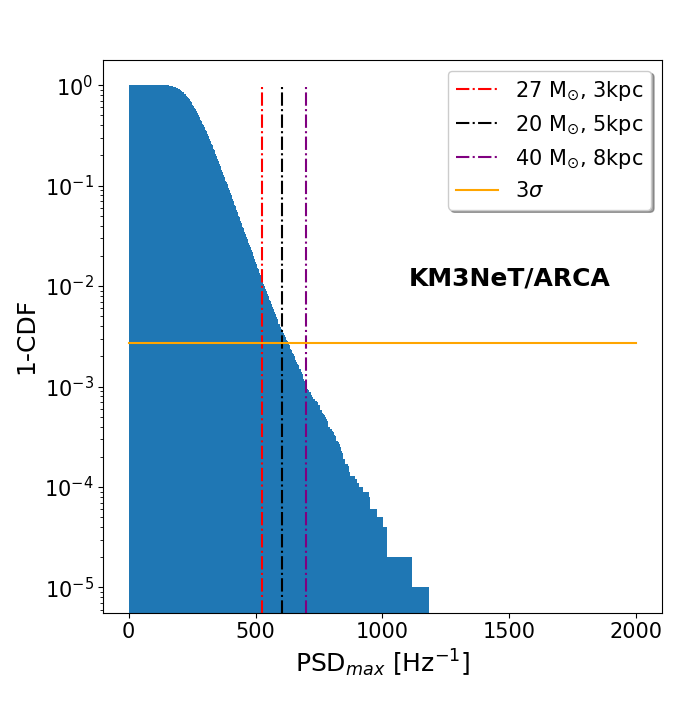}%
    \includegraphics[width=0.45\textwidth]{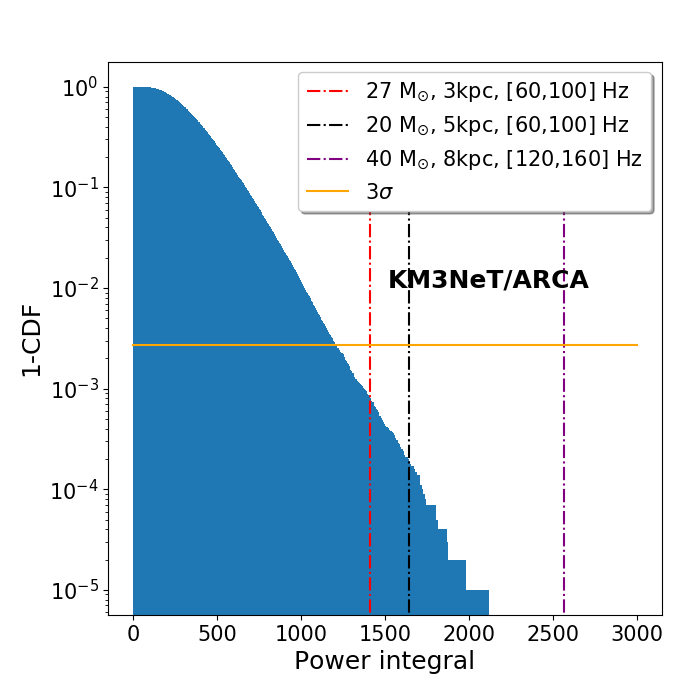}%
    \caption{Background anti-cumulative density function built from ARCA pseudo-experiments (blue distribution). On the left for the maximum power (model independent approach). On the right for the power integral around the SASI frequency predicted by the model. The horizontal line (orange) indicates the $3\,\sigma$ value. The vertical lines, dashed red, black and purple correspond to the expectation, given by the average of pseudo-experiments, for the signal plus background scenario for the $\SI{27}{\solarmass}$, $\SI{20}{\solarmass}$ and $\SI{40}{\solarmass}$ progenitors at 3, 5 and 8 kpc, respectively.}
    \label{fig:PDFs}
\end{figure*}

The probability density functions, built as the anti-cumulative density function (1-CDF), of PSD$_{max}$ and of the power integral for the pure-background scenario are estimated with pseudo-experiments. The corresponding anti-cumulative distributions are shown in Figure~\ref{fig:PDFs} for the ARCA detector. These distributions are compared to the expectation for the signal plus background scenario, evaluated as the average of the pseudo-experiments (dashed vertical lines), to infer the significance from the corresponding p-value.

 The ORCA and ARCA detected neutrino signals will have a relative delay due to their different locations at Earth. Due to the uncertainty on the source direction, the successful synchronisation of the light curve data at sub-ms precision for a combined analysis is not guaranteed. Here, the analysis is applied to the detected light curve independently for each detector, combining the significances according to the Equation~\ref{eq:combination}. 

 The obtained sensitivities are summarised in Table~\ref{t:sasi}. For the two CCSN progenitors of $\SI{20}{\solarmass}$ and $\SI{27}{\solarmass}$, the results are provided for the distance at which the model independent approach reaches a sensitivity close to $3\,\sigma$. For the black hole forming scenario ($\SI{40}{\solarmass}$), they are given for a source at the Galactic Centre. The same systematic uncertainties described in Section~\ref{s:t0} are taken into account, with the exception of bioluminescence, here evaluated separately. Combining ORCA and ARCA, the variability of the optical noise due to bioluminescence yields an additional $\pm$ $0.3\,\sigma$ uncertainty in the SASI sensitivity results.

For comparison, the IceCube and Hyper-Kamiokande detectors will be sensitive to SASI oscillations for the cases of the $\SI{20}{\solarmass}$ and $\SI{27}{\solarmass}$ up to a distance of $\sim\SI{20}{kpc}$~\cite{Tamborra:2013laa-SASI,Migenda:2016xnc}. For the more massive progenitor with $\SI{40}{\solarmass}$~\cite{Walk:2019miz-3DBH}, their SASI detection capabilities go as far as $\sim\SI{250}{kpc}$.

\section{Conclusions}
\label{s:conclusion}

An analysis method for the observation of $\sim\SI{10}{MeV}$ core-collapse supernova neutrinos in KM3NeT has been established. It is based on the detection of an excess of hit coincidences above the optical backgrounds that are produced by radioactive decays in seawater, bioluminescence and atmospheric muons. The multi-PMT design of the KM3NeT DOM is instrumental to this method. Data from the first six deployed detection units of KM3NeT in the ARCA and ORCA sites have been analysed to study and characterise the background features. The signal expectation for a CCSN neutrino emission in KM3NeT is evaluated considering four different CCSN flux models and a detailed simulation of the detector response. The optical backgrounds are suppressed by dedicated filtering methods. An event selection based on the number of hit PMTs in a coincidence has been optimised to maximise the the distance at which a $5\,\sigma$ discovery is achieved. Combining ARCA and ORCA sensitivities, KM3NeT will be able to detect the next Galactic explosion with a $5\,\sigma$ discovery potential. For the considered black-hole forming scenario, the sensitivity extends well beyond the Large Magellanic Cloud.

The astrophysical potential of a CCSN detection in KM3NeT has been evaluated, including the case of a black-hole forming collapse. For a supernova at 10 kpc, KM3NeT will be able to estimate the mean energy of the CCSN neutrinos with sub-MeV accuracy, with previous knowledge of the other spectral parameters. The time of arrival of the neutrino burst can be estimated with an uncertainty as low as $\SI{3}{\milli\second}$ for a supernova at $\SI{5}{kpc}$ ($\SI{7.5}{\milli\second}$ for the most conservative flux assumption, at the same distance). A $3\,\sigma$ sensitivity to the signature of the \emph{standing accretion shock instability} (SASI) is reached for Galactic progenitors at distances between 3 ($\SI{27}{\solarmass}$) and 5~kpc ($\SI{20}{\solarmass}$), using the model independent search. In the black-hole forming scenario, the SASI is detectable beyond the Galactic Centre.

For a Galactic CCSN, besides the precise estimation of the arrival time of the burst, KM3NeT will be able to promptly share the data of the neutrino light curve with millisecond time resolution. These two key parameters are crucial for multi-messenger networks as SNEWS2.0 to confirm the detection and provide an early and precise localisation of the source to the astronomy community.

\begin{acknowledgements}
The authors acknowledge the financial support of the funding agencies:
Agence Nationale de la Recherche (contract ANR-15-CE31-0020),
Centre National de la Recherche Scientifique (CNRS), 
Commission Europ\'eenne (FEDER fund and Marie Curie Program),
Institut Universitaire de France (IUF),
LabEx UnivEarthS (ANR-10-LABX-0023 and ANR-18-IDEX-0001),
Paris \^Ile-de-France Region,
France;
Shota Rustaveli National Science Foundation of Georgia (SRNSFG, FR-18-1268),
Georgia;
Deutsche Forschungsgemeinschaft (DFG),
Germany;
The General Secretariat of Research and Technology (GSRT),
Greece;
Istituto Nazionale di Fisica Nucleare (INFN),
Ministero dell'Universit\`a e della Ricerca (MIUR),
PRIN 2017 program (Grant NAT-NET \\ 2017W4HA7S)
Italy;
Ministry of Higher Education Scientific Research and Professional Training,
ICTP through Grant AF-13,
Morocco;
Nederlandse organisatie voor Wetenschappelijk Onderzoek (NWO),
the Netherlands;
The National Science Centre, Poland (2015/18/E/ST2/00758);
National Authority for Scientific Research (ANCS),
Romania;
Ministerio de Ciencia, Innovaci\'{o}n, Investigaci\'{o}n y Universidades (MCIU): Programa Estatal de Generaci\'{o}n de Conocimiento (refs. PGC2018-096663-B-C41, -A-C42, -B-C43, -B-C44) (MCIU/FEDER), Severo Ochoa Centre of Excellence and MultiDark Consolider (MCIU), Junta de Andaluc\'{i}a (ref. SOMM17/6104/UGR), Generalitat Valenciana: Grisol\'{i}a (ref. GRISOLIA/2018/119) and GenT (ref. CIDEGENT/2018/034 and CIDEGENT/2019/043) programs, La Caixa Foundation (ref. LCF/BQ/IN17/11620019), EU: MSC program (ref. 713673),
Spain.
This work has also received funding from the European Union’s Horizon 2020 research and innovation program under grant agreement No 739560.
\end{acknowledgements}

\bibliographystyle{hunsrt} 
\bibliography{references}

\end{document}